\documentclass[aps,prx,reprint,showpacs,superscriptaddress,floatfix,nofootinbib]{revtex4-2}
\usepackage{amsmath,amssymb,amsfonts,mathtools}
\usepackage{graphicx}
\usepackage{xcolor}
\usepackage[colorlinks=true,allcolors=blue]{hyperref}
\usepackage[T1]{fontenc}
\usepackage{lmodern}
\usepackage{amsmath}
\usepackage{bm}
\usepackage[normalem]{ulem}
\usepackage{graphicx}
\usepackage{amsmath}
\usepackage{xcolor}
\usepackage{booktabs}
\usepackage{multirow}
\usepackage{wasysym}
\usepackage{slashed}
\usepackage{ragged2e}
\usepackage{comment}
\usepackage{xspace}
\usepackage{mathrsfs}
\usepackage{orcidlink}
\usepackage{listings}
\usepackage{amsthm}
\usepackage{tikz}
\usetikzlibrary{arrows,arrows.meta}
\usepackage[braket,qm]{qcircuit}
\usepackage[normalem]{ulem}
\definecolor{forestgreen}{rgb}{0.13, 0.55, 0.13}
\definecolor{green}{RGB}{34,139,34}

\newcommand{\Rea}{\operatorname{Re}}
\newcommand{\diag}{\operatorname{diag}}
\newcommand{\PP}{\mathcal P}
\newcommand{\QQ}{\mathcal Q}
\newcommand{\FF}{\mathcal F}
\newcommand{\K}{\mathsf K}
\newcommand{\DD}{\mathsf D}
\newcommand{\one}{\mathbf1}
\newtheorem{theorem}{Theorem}
\newtheorem{corollary}{Corollary}
\newtheorem{lemma}{Lemma}

\usepackage{lmodern,microtype}
\newcommand{\Tr}{\operatorname{Tr}}

\newcommand{\CHANGE}[1]{\textcolor{black}{#1}}

\makeatletter
\renewcommand{\onecolumngrid}{%
  \do@columngrid{one}{\@ne}%
  \let\set@footnotewidth\set@footnotewidth@one
  \let\compose@footnotes\compose@footnotes@one
  \onecolumn@grid@setup
}
\makeatother

\newcommand{\jmax}{j_{\mathrm{max}}}

\definecolor{forestgreen}{rgb}{0.13, 0.55, 0.13}

\definecolor{green}{RGB}{34,139,34}

\newcommand{\SUtwo}{\mathrm{SU}(2)}
\newcommand{\hc}{\mathrm{H.c.}}
\newcommand{\Ree}{\operatorname{Re}}
\newcommand{\epsj}{\epsilon_{\bm j}}
\newcommand{\mux}{\bm\mu_x}
\newcommand{\muy}{\bm\mu_y}
\newcommand{\bj}{\bm j}
\newcommand{\bk}{\bm k}

\newcommand{\half}{\tfrac{1}{2}}

\begin{document}

\title{Non-stabilizerness and entanglement in $(2+1)$-dimensional $\mathrm{SU}(2)$ lattice gauge theory using tensor networks}

\author{Raghav~G.~Jha\orcidlink{0000-0003-2933-0102}}
\email{raghav.govind.jha@gmail.com}
\affiliation{Department of Physics and Astronomy, North Carolina State University, Raleigh, North Carolina 27695, USA}

\author{Jaber~I.~Taher\orcidlink{0009-0003-0189-4317}}
\email{mtaher@ncsu.edu}
\affiliation{Department of Physics and Astronomy, North Carolina State University, Raleigh, North Carolina 27695, USA}

\author{Muhammad Asaduzzaman\orcidlink{0000-0001-7559-3873}}
\affiliation{Department of Physics and Astronomy, North Carolina State University, Raleigh, North Carolina 27695, USA}

\author{Goksu~C.~Toga\orcidlink{0000-0002-0316-2502}}
\affiliation{Department of Physics and Astronomy, North Carolina State University, Raleigh, North Carolina 27695, USA}

\author{Bojko N. Bakalov\orcidlink{0000-0003-4630-6120}}
\affiliation{Department of Mathematics, North Carolina State University, Raleigh, North Carolina 27695, USA}

\author{Alexander~F.~Kemper\orcidlink{0000-0002-5426-5181}}
\affiliation{Department of Physics and Astronomy, North Carolina State University, Raleigh, North Carolina 27695, USA}

\begin{abstract}
We study non-stabilizerness (magic) in the ground state of $(2+1)$-dimensional $\mathrm{SU}(2)$ Hamiltonian lattice gauge theory with matter, formulated in the dressed-site basis in the hardcore-gluon truncation and restricted to the zero baryon-number sector. Using matrix product states, we compute three facets of magic: the second-order stabilizer R\'enyi entropy (SRE) $M_2$, its non-local component $M_2^{\rm NL}$, estimated by the SRE $M_2(\tilde\psi)$ of the computational-basis Schmidt representative $\tilde\psi$, and a lower bound in terms of the anti-flatness $F$ of the entanglement spectrum. We also prove a stronger form of the sandwich relation: $-\log_2(1-4F)\le M_2^{\rm NL}\le M_2$; the lower bound rests on a stronger inequality that we obtain for arbitrary Schmidt bases and rank, with an optimal constant. Combined with recent results, this yields the unconditional bracket $\max\{-\log_2(1-4F),\,M_2(\tilde\psi)-4\}\le M_2^{\rm NL}\le M_2(\tilde\psi)$. We emphasize a structural distinction that makes the non-local quantities the physically preferred diagnostics: whereas the full SRE depends on the (non-unique) encoding of the gauge-invariant local Hilbert space into qubits, the non-local magic, its estimate $M_2(\tilde\psi)$, and the anti-flatness are invariant under site-local re-encodings and are therefore intrinsic to the state and bipartition. By varying the gauge coupling for two values of the fermion mass on lattices up to $8\times 8$ ($\sim 288$ qubits) with bond dimension up to $128$, we find that the non-local magic, as estimated by $M_2(\tilde\psi)$, furnishes a sharper and more bond-dimension-friendly probe of the gauge-matter delocalization crossover compared to the full SRE or the gauge-invariant entanglement entropy, retaining a clear signal at bond dimensions well below those needed to converge the ground state itself.
\end{abstract}

\maketitle

\section{Introduction}

An important and profound application of fault-tolerant quantum computers is the
simulation of quantum field theories, which describe the fundamental
interactions of nature. A central question in this program is what quantum
resources are required to represent and evolve the states of a gauge theory, and
in particular how those resources grow as the continuum limit is approached.
Non-stabilizerness, or magic, has recently emerged as a sharp diagnostic of this
cost, complementary to entanglement; yet numerical studies of magic in non-Abelian gauge
theories have so far been limited to one (spatial) dimension~\cite{Santra:2025dsm,Ebner:2025pdm,Jha:2026ror}. In this work, we move to $(2+1)$ dimensions, the smallest dimension in which genuine magnetic (plaquette) dynamics and a gauge–matter delocalization crossover arise, and we study both local and non-local magic of the ground state of $\SUtwo$ lattice gauge theory with matter using tensor networks in the hardcore gluon approximation~\cite{Cataldi:2023xki}. The name ``hardcore'' refers to the fact that the only retained states of the gauge field are ones generated from the bare vacuum with at most single application of the parallel transporter operator.

The Standard Model of particle physics is formulated in terms of quantum fields,
but extracting quantitative predictions, for example, in the strongly coupled
regime of the strong interaction becomes difficult where perturbation theory
fails and one must resort to numerical methods. Lattice gauge theory provides a
non-perturbative framework by introducing a finite lattice spacing, with the
continuum theory recovered as the spacing is taken to zero at fixed physical
scales. In the Euclidean formulation, lattice calculations have achieved
remarkable success in describing the non-perturbative properties of quantum chromodynamics (QCD),
including hadron spectroscopy and hadronic
structure~\cite{BMW:2008jgk,FlavourLatticeAveragingGroup:2019iem}. However, these methods
encounter the sign problem when the Euclidean action becomes complex, as
occurs at finite baryon density or in the presence of certain topological
terms~\cite{Troyer:2004ge,Gattringer:2016kco,Unsal:2012zj}: the weight can no
longer be interpreted as a probability distribution, obstructing the direct use
of importance-sampling Markov chain Monte Carlo
simulations~\cite{Jha:2021exo}. Real-time observables are equally out of reach,
since analytically continuing imaginary-time correlators to real-time is an ill-posed problem,
leaving hadronic dynamics largely inaccessible to traditional sampling-based
methods~\cite{Zohar:2021nyc}.

The Hamiltonian formulation circumvents both obstacles: it is free of the sign
problem and provides direct access to real-time dynamics. However, it brings its own
challenges, as the gauge field on each link spans an infinite-dimensional Hilbert
space that must be truncated, the entanglement of the relevant states grows
rapidly with system size, and imposing gauge invariance is considerably more
delicate than in the Lagrangian approach. In one and two spatial dimensions, classical methods based on tensor networks, variational Monte Carlo, and neural quantum states have made substantial progress and are now probing small systems even in higher dimensions~\cite{Spriggs:2025sea, Rouxinol:2026eur, Rouxinol:2026vjl,Zwolfer:2026yuc}. 

Quantum computers offer a promising platform for studying Hamiltonian lattice
gauge theories~\cite{Banuls:2019bmf,Byrnes:2005qx} in regimes beyond the reach of
classical methods. As we enter the early fault-tolerant era of quantum
simulators, it is timely to ask what quantum resources are needed to represent
and simulate the ground states of lattice gauge theories, and how these costs
compare with classical approaches. 

Entanglement alone does not settle this
question: stabilizer states can be highly entangled yet remain efficiently
classically simulable. Non-stabilizerness, or magic~\cite{BravyiKitaev:2005,
Veitch:2014}, is the complementary resource, relevant both to stabilizer-based
classical simulation and to fault-tolerant quantum computation.

Magic has recently been studied across a range of Abelian and non-Abelian gauge
theories~\cite{Falcao:2024msg,Ebner:2025pdm,Santra:2025dsm,
Grieninger:2026bdq,Esposito:2024uzw,Jha:2026ror,Froland:2026aff},
as well as in gluon scattering~\cite{Gargalionis:2025iqs, Chu:2026yxm} and dense
neutrino gases~\cite{Hite:2026fsj}. On the theoretical side, arguments based on
von Neumann algebras show that the physical states of a quantum field
theory, which generate a Type~III algebra, are states of extremely high
magic~\cite{Benedetti:2026mfy, Moosa:2026nqz}, implying that quantum simulation
of lattice gauge theories will demand substantial quantum resources as the
continuum limit is approached. The same trend was seen numerically in our earlier
study of $1+1$-dimensional $\SUtwo$ gauge theory, where approaching the continuum
pushed the ground state into a regime of increasingly large magic~\cite{Jha:2026ror}.
These results, established both analytically and numerically, signal the
difficulty of studying lattice gauge theories on a 
quantum computer in the early fault tolerant era. 

In this work, we pursue the study of non-stabilizerness (magic) in the ground state of (2+1)-dimensional $\SUtwo$ lattice gauge theory with matter in the hardcore limit of dressed-site formulation using an approach reminiscent of quantum link and gauge magnets~\cite{Horn:1981kk, Orland:1989st, Chandrasekharan:1996ih, Tagliacozzo:2012df, Cataldi:2023xki} in the zero baryon-number sector. 
We probe three different facets of quantum magic: extensive second-order stabilizer Rényi entropy (SRE), non-local SRE, and the lower bound of non-local SRE in terms of anti-flatness by using a stronger inequality than considered in earlier works~\cite{Cao:2024nrx}. \CHANGE{We also relate this bound to recent exact results on non-local magic~\cite{Sierant:2026qji, Sierant:2026lvg}: the anti-flatness bound never exceeds $\log_2(64/37)\approx0.79$ bits and is the only non-trivial lower bound when $M_2^{\rm NL}<4$ bits, which covers all of our data, whereas the spectral bound of Ref.~\cite{Sierant:2026lvg} takes over when the non-local magic grows large.} We investigate these three magic-related observables, and other gauge-invariant observables for different values of the coupling $g$, and locate the signatures of the gauge–matter delocalization crossover at $g_{\star}$~\cite{Cataldi:2023xki, Xu:2025ean, Rouxinol:2026vjl}. This crossover marks the change of behavior from localized state at large $g^2$ to flux ordered/delocalized as $g^2 \to 0$. 

\begin{figure}
    \centering
    \includegraphics[width=1.01\linewidth]{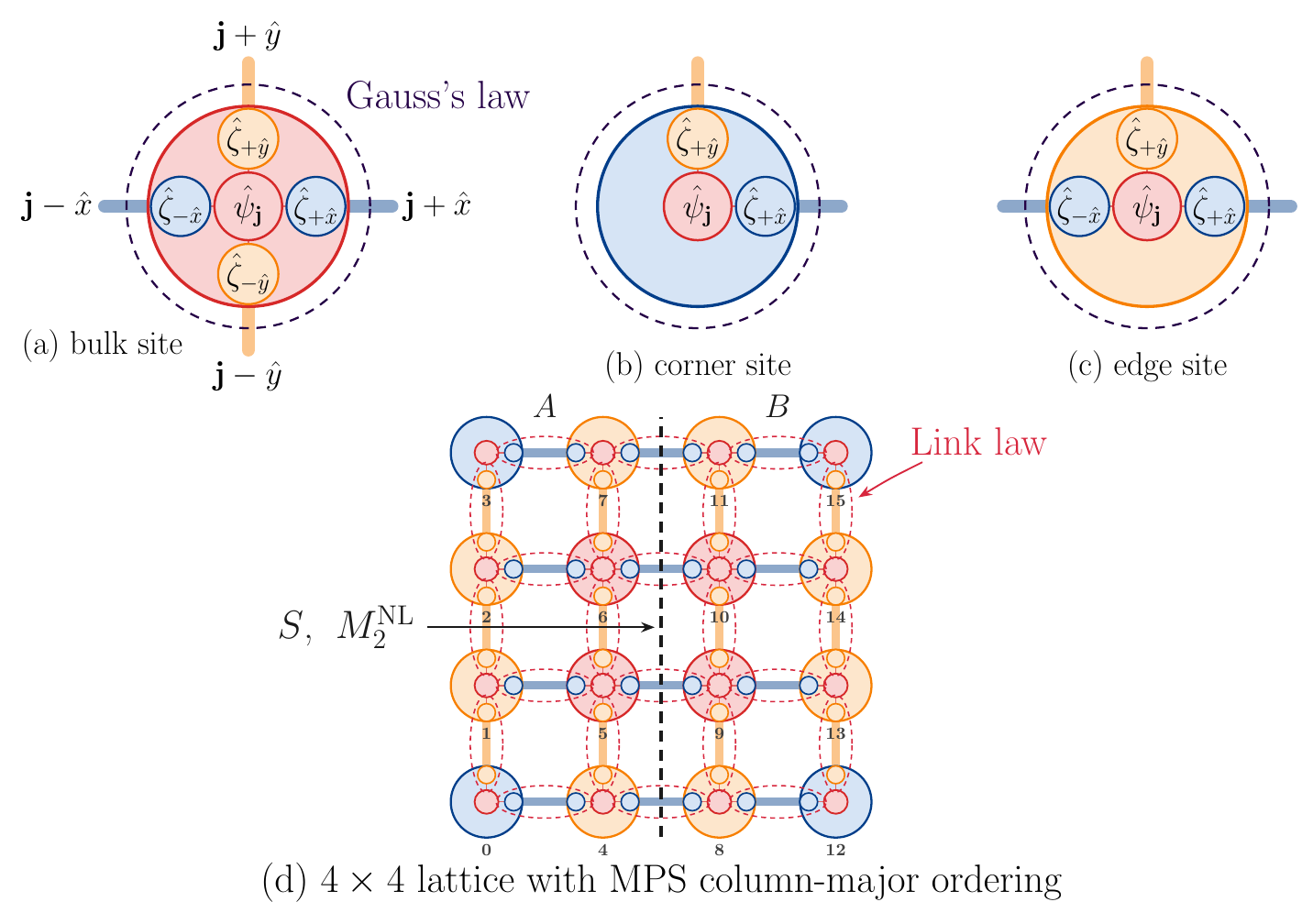}
    \caption{Single dressed-site composed of matter site and semi-links shown for a two-dimensional lattice where we have three types of dressed sites --- bulk, edge, and corner corresponding to four, three, and two semi-links/rishons (shown at top). An example of $4 \times 4$ lattice using these dressed sites and the bipartition (black dashed) we use for computation of entropy and magic observables. The site indices in the column-major pattern are also shown. See Appendix~\ref{app:sec1} for additional details.}
    \label{fig:dressed_site}
\end{figure}

\section{Hamiltonian formulation in the dressed basis}

We consider $\SUtwo$ gauge theory in $2+1$ dimensions with matter on a two-dimensional square lattice, using the standard Hamiltonian formulation of Kogut and Susskind \cite{Kogut:1974ag}. In the conventional formulation, the gauge field resides on the links. We instead use the dressed-site, or semi-link/rishon representation, in which the gauge degrees of freedom are absorbed into the sites by explicitly solving Gauss's law with a truncation on the allowed values for the irreducible representations of the gauge group. This leads to a finite-dimensional local Hilbert space, with interactions between nearest neighbor and around plaquettes. We show the lattice setup in Fig.~\ref{fig:dressed_site}.  

The construction proceeds in two steps. First, the gauge field on each link is truncated to the two lowest irreducible representations of $\SUtwo$, with $j=0$ and $j=1/2$ --- the hardcore-gluon truncation. Since the color-electric energy of an irreducible representation grows as $g^2 j(j+1)/2$, higher spin representations are energetically suppressed, making this truncation well-controlled at strong and intermediate couplings, i.e., for $g \gg 1$ in Eq.~\eqref{eq:Hpuredim}. 
Recovering the exact continuum theory requires the representation cutoff to increase, i.e., $j_{\max}\to\infty$ while keeping physical mass scale fixed and taking $a\to0$. We do not attempt this in this work. However, it is an interesting question of how large this truncation has to be to retrieve the continuum physics and whether there is some assistance provided by gauge invariance and universality in this context. For example, in one dimension, we found that 
one can obtain an accurate central charge even with $j_\text{max}=1/2$~\cite{Jha:2026ror}. 
We leave these questions regarding higher dimensions for future work. 

Once the suitable truncation has been fixed, each truncated link is decomposed into a pair of semi-links or rishon modes, to the left and right of each link. A semi-link is an auxiliary color-doublet fermionic mode with local Hilbert space: 
\begin{equation}
  \mathcal{H}_{\rm rishon}
  =
  {\rm span}\big\{\ket{0},\ket{\textcolor{red}{r}},\ket{\textcolor{forestgreen}{g}}\big\},
  \qquad
  \dim\mathcal{H}_{\rm rishon}=3,
\end{equation}
where double occupancy is excluded by the truncation. The rishons are then absorbed into the lattice sites together with the staggered matter fields.

The resulting Hamiltonian can then be expressed in terms of on-site operators, nearest-neighbor hopping, and plaquette interactions. The connection between the two rishons belonging to the same link is obtained by requiring them to carry the same $\SUtwo$ flux,
\begin{equation}
  \hat T^{2}_{\rm head}(\mathbf{j})
  =
  \hat T^{2}_{\rm tail}(\mathbf{j^\prime}),
  \qquad
  \langle\mathbf j,\mathbf j^{\prime}\rangle\ \text{nearest neighbors}.
  \label{eq:linkconstraint}
\end{equation}
The main simplification of the dressed-site construction is that Gauss's law becomes a strictly \emph{on-site} constraint. Physical states are color singlets under the combined $\SUtwo$ generator of matter and rishons:
\begin{equation}
  \hat G^{2}_{\mathbf j}\ket{\rm phys}=0,
  \qquad
  \hat G^{a}_{\mathbf j}
  =
  \hat S^{a}_{\rm matter}(\mathbf j)
  +
  \sum_{\ell}\hat T^{a}_{\ell}(\mathbf j).
  \label{eq:gauss}
\end{equation}
where $\hat{G}^2_\mathbf j = \sum_{a=1}^{3}\hat{G}^a_\mathbf j\cdot \hat{G}^a_\mathbf j$ and $
  \hat S^{a}_{\rm matter}(\mathbf j)
   =\frac12\sum_{\alpha\beta}\psi^{\dagger}_{\mathbf j,\alpha}\,
     \sigma^{a}_{\alpha\beta}\,\psi_{\mathbf j,\beta}$ for 
  $a=1,2,3$ and $\alpha, \beta \in \{\textcolor{red}{r}, \textcolor{green}{g}\}$.
A site touches four links in the bulk, three on an edge, and two at a corner. The corresponding raw on-site Hilbert-space dimensions are therefore
$4\times3^{n_{\rm links}}=324$, $108$, and $36$ for bulk, edge, and corner, respectively. These spaces
contain many states that violate Gauss's law. Imposing the gauge singlet condition reduces the local dimensions to: 
\begin{equation}
  \dim\mathcal H^{\rm phys}_{\rm site}
  =
  30\;(\text{bulk}),\qquad
  13\;(\text{edge}),\qquad
  6\;(\text{corner}).
  \label{eq:dims}
\end{equation}
Because Eq.~\eqref{eq:gauss} is local, this projection is performed once when constructing the on-site basis. The Hamiltonian then acts entirely within the
physical space, so no further gauge projection is required. The link constraint~\eqref{eq:linkconstraint}, which couples neighboring sites, is enforced separately.

Each corner, edge, and bulk dressed-site requires 3, 4, and 5 qubits, respectively, leading to a total count of $N_q
  =
  3N_{\rm corner}
  +4N_{\rm edge}
  +5N_{\rm bulk}
  =
  5L_xL_y-2L_x-2L_y$ qubits. Additional details can be found in Appendix~\ref{app:gauge-singlet}. With the conventions of Ref.~\cite{Cataldi:2023xki}, and setting $c=\hbar=1$, the Hamiltonian with which we are working is: 
\begin{align}
 H
 ={}& \frac{1}{2a}
 \sum_{\bj\in\Lambda}\sum_{\alpha,\beta}
 \Big[
 -i\,\psi_{\bj,\alpha}^{\dagger}
 U^{\alpha\beta}_{\bj,\bj+\mux}
 \psi^{\phantom{\dagger}}_{\bj+\mux,\beta}
 \nonumber\\
 &\hspace{0.5cm}
 -\epsj\,
 \psi_{\bj,\alpha}^{\dagger}
 U^{\alpha\beta}_{\bj,\bj+\muy}
 \psi^{\phantom{\dagger}}_{\bj+\muy,\beta}
 +\hc
 \Big] 
 \nonumber\\
 &+m_0\sum_{\bj}\epsj
 \sum_{\alpha}\psi_{\bj,\alpha}^{\dagger}\psi^{\phantom{\dagger}}_{\bj,\alpha}
 +\hat H_{\mathrm{pure}},
 \label{eq:Hdim_main}
\end{align}
with 
\begin{align}
 \hat H_{\mathrm{pure}}
 ={}& \frac{g^2}{2a}
 \sum_{\bj}\left(
 \hat E^2_{\bj,\bj+\mux}
 +\hat E^2_{\bj,\bj+\muy}
 \right) -\frac{1}{ag^2}
 \sum_{\square}
 \Ree\Tr \,\Box, 
 \label{eq:Hpuredim}
\end{align}
where we have used the shorthand notation $\Box = U_{\bj,\bj+\mux}
 U_{\bj+\mux,\bj+\mux+\muy}
 U_{\bj+\muy,\bj+\mux+\muy}^{\dagger}
 U_{\bj,\bj+\muy}^{\dagger}$,   
 $\epsj=(-1)^{j_x+j_y}$ is the staggered phase factor,
 and $\alpha, \beta$ are $\SUtwo$ indices 
 $\in \{\textcolor{red}{r}, \textcolor{green}{g}\}$ and $a$ is the lattice spacing.
 The terms in Eq.~\eqref{eq:Hdim_main} describe gauge-covariant hopping in two spatial dimensions, the staggered mass, the electric energy, and the magnetic energy, respectively. 
 In dimensionless units, $\hat H\to a\hat H$ and $m=m_0a$.
 In the staggered formulation, a filled site on the odd sublattice ($\epsilon_{\bm j}=-1$) represents the occupied Dirac sea and an empty site on the even sublattice ($\epsilon_{\bm j}=+1$) the empty one, so $b$ counts quarks minus anti-quarks and is given by:
\begin{equation} b = \frac{1}{2} \sum_{\bm j}\left(\sum_{\alpha = \textcolor{red}{r}, \textcolor{green}{g}}\psi^{\dagger}_{\bm j,\alpha}\psi_{\bm j,\alpha} -\bigl(1-\epsilon_{\bm j}\bigr)\right).
\label{eq:baryon} \end{equation}
The $b=0$ sector therefore has zero net baryon density and fixes the total fermion number to $N_{\rm tot}=\sum_{\bm j,\alpha}\psi^{\dagger}_{\bm j,\alpha}\psi_{\bm j,\alpha}=n_{\rm sites}$. Throughout the paper, we work with open boundary conditions and in the zero baryon number sector, $b=0$.

\section{Magic, entanglement, and other observables} 

Quantumness is not captured by any single resource --- different measures quantify different aspects of what makes a quantum state hard to describe or simulate classically. Two of the most widely studied are entanglement entropy and
non-stabilizerness (or magic), which probe fundamentally different features of a
quantum state.

Non-stabilizerness characterizes how far a state lies from the stabilizer
polytope, the convex hull of the set of stabilizer states. Stabilizer states,
together with Clifford operations and Pauli measurements form a class of quantum states
that can be simulated efficiently (in polynomial time) 
on a classical computer by virtue of the Gottesman--Knill theorem~\cite{Gottesman:1998hu}. In the Pauli basis, a pure
$n$-qubit stabilizer state has a particularly simple structure: of the $4^n$
Pauli strings, only $2^n$ have nonzero expectation values, each with
$\langle P\rangle=\pm1$. Magic can therefore be viewed as the spreading of Pauli
weight beyond this minimal stabilizer structure.

A convenient way to quantify this spreading is through the stabilizer R\'enyi
entropies (SRE), introduced by Leone, Oliviero, and Hamma~\cite{Leone:2021rzd}. For R\'enyi index $\alpha=2$, the SRE of a pure state
$\ket{\psi}$ of $n$ qubits is
\begin{equation}
M_2(\ket{\psi})
=
-\log_2\!\left[
\frac{1}{2^n}
\sum_{P \in \mathcal{P}}
\langle\psi|P|\psi\rangle^4
\right],
\label{eq:M2}
\end{equation}
where the sum runs over all $n$-qubit Pauli strings in the Pauli group $\mathcal{P}$. The quantity $M_2$ vanishes for stabilizer states and grows as the Pauli weight becomes more
broadly distributed. Beyond its role as a measure of non-stabilizerness, $M_2$
is experimentally accessible through a four-copy protocol, analogous to the
two-copy SWAP protocol~\cite{Ekert:2002, Daley:2012} used to measure the second
R\'enyi entanglement entropy. Because the summation in Eq.~\eqref{eq:M2}
contains $4^n$ terms, a direct evaluation is not possible beyond a few qubits.
We evaluate the full-state $M_2$ by Pauli sampling of the
converged MPS, following Ref.~\cite{Lami:2023naw}. This is feasible for the smaller lattices; for the larger systems, we report only \CHANGE{the non-local quantities, $M_2(|\tilde\psi\rangle)$ (defined below) and the anti-flatness,} which do not require Pauli sampling and, as we show, already capture the crossover.

\paragraph{Non-local magic.}
The full SRE mixes two physically distinct contributions: magic that can be
created or removed by acting within each subsystem independently, and magic that
is tied to the correlations across a bipartition. The latter is isolated by the
\emph{non-local} magic~\cite{Cao:2024nrx, Sierant:2026qji}. For a bipartition
$\mathcal{H}=\mathcal{H}_A\otimes\mathcal{H}_B$ and a pure state $|\psi\rangle$, we define
\begin{equation}\label{eq:MNL}
M_n^{\mathrm{NL}}(|\psi\rangle)
= \min_{U_A\otimes U_B}\,
   M_n\bigl((U_A\otimes U_B)|\psi\rangle\bigr),
\end{equation}
where the minimization runs over all products of local unitaries $U_A$ and
$U_B$ acting on $\mathcal{H}_A$ and $\mathcal{H}_B$. 
Note that, by definition,
\begin{equation}\label{eq:MNLM}
M_n^{\mathrm{NL}}(|\psi\rangle)
\le
M_n(|\psi\rangle).
\end{equation}
Minimizing over local unitaries removes exactly the
magic that can be disentangled by local basis choices and retains only the piece
associated with the bipartite correlations; local unitaries can neither generate
nor erase it, so $M_n^{\mathrm{NL}}$ is a property of the entanglement structure
rather than of the local bases. The exact minimization in Eq.~\eqref{eq:MNL} is
intractable in general. In practice, we \CHANGE{estimate} the non-local magic from the Schmidt reference state of
the central cut following Refs.~\cite{Liu:2026nah, Cao:2024nrx}: we retain the
leading $\chi$ Schmidt coefficients $\sqrt{\lambda_i}$ of the converged MPS, form
the reference state
$ |\tilde{\psi}\rangle = \sum_i \sqrt{\lambda_i}\,|i\rangle_A|i\rangle_B$,
where $|i\rangle_A$ and $|i\rangle_B$ represent computational basis states corresponding to subsystems $A$ and $B$.
\CHANGE{Its SRE, $M_2(|\tilde\psi\rangle)$, is an unconditional upper bound on $M_2^{\rm NL}$ (Corollary~\ref{cor:bracket}); it coincides with $M_2^{\rm NL}$ under the conjecture of Ref.~\cite{Liu:2026nah} that the descending diagonal labelling is optimal, which has been proven for Schmidt rank $\chi\le6$ and for dyadic-staircase spectra, and it exceeds $M_2^{\rm NL}$ by at most $4$ bits in general~\cite{Sierant:2026lvg}. We use $M_2(|\tilde\psi\rangle)$ as our estimate of $M_2^{\rm NL}$ and show it, labelled as such, in all figures; at the central cut of our MPS we retain the full Schmidt spectrum (rank $\le\chi$).}
Additional details are provided in Appendix~\ref{app:subsec_lowerbound_and_nlmagic}.

\paragraph{Anti-flatness and the sandwich relation.}
We also use the anti-flatness, which lower-bounds the non-local magic. For a bipartition with reduced density matrix $\hat\rho_A$,
the anti-flatness defined by
\begin{equation}
  F
  =
  \operatorname{Tr}\hat\rho_A^{3}
  -
  \bigl(\operatorname{Tr}\hat\rho_A^{2}\bigr)^{2},
  \label{eq:anti-flatness-def}
\end{equation}
measures how far the entanglement spectrum deviates from a flat distribution:
$F$ vanishes for stabilizer states, whose entanglement spectra are flat across
any bipartition. Since $F$ depends only on the Schmidt spectrum, it is invariant
under local unitaries on either side of the cut, and it bounds the non-local
magic from below. The most common form of inequality
used is the sandwich relation
$4F \;\le\; M_2^{\rm NL} \;\le\; M_2$. In a recent work, Liu and Cui~\cite{Liu:2026twg} obtain
$M_2^{\rm NL}(|\tilde\psi\rangle)\ge-\log_2(1-4F)$ for the ordered computational-basis Schmidt representative $|\tilde\psi\rangle$.
The global optimality was proved by Sierant~\cite{Sierant:2026lvg}
for dyadic-staircase spectra and for Schmidt rank of at most six. 
The general case has been open, and we prove the inequality 
for arbitrary Schmidt bases and rank.
 
\begin{theorem}\label{thm:sandwich}
Let $|\psi\rangle$ be a normalized pure state on two qubit registers $A,B$ of
dimensions $d_A=2^{n_A}$ and $d_B=2^{n_B}$, and let
$\rho_A=\Tr_B|\psi\rangle\langle\psi|$ be its reduced density matrix. Define the stabilizer purity
\[
 W=\frac1{d_Ad_B}\sum_{P,Q}\langle\psi|P\otimes Q|\psi\rangle^4 ,
\]
where $P$ and $Q$ run over the $d_A^2$ and $d_B^2$ tensor products of
$\{\mathbb{I}, X, Y, Z\}$ on their respective registers, so that
$P\otimes Q$ runs over the Hermitian Pauli strings of the joint system. In
terms of $W$, the linear and R\'enyi-2 stabilizer entropies are
$M_{\rm lin}=1-W$ and $M_2=-\log_2W$, and
$M_{\rm lin}^{\rm NL},M_2^{\rm NL}$ denote their non-local components,
obtained by minimizing $M_{\rm lin},M_2$ over local unitaries
$U_A\otimes U_B$. Then
\begin{equation}
 4F\ \le\ M_{\rm lin}^{\rm NL},
 \qquad
 -\log_2(1-4F)\ \le\ M_2^{\rm NL}\ \le\ M_2 .   
 \label{eq:strong_inequality}
\end{equation}
The lower bounds hold for arbitrary Schmidt rank and basis, without assuming the computational-basis Schmidt representative is optimal, and the logarithm
is finite since $W\ge(d_Ad_B)^{-1}>0$.
\end{theorem}
The proof is presented in Appendix~\ref{app:proof_of_tighter_bound}.
\CHANGE{In addition, Sierant has shown~\footnote{We thank Piotr Sierant for helpful correspondence regarding this after the first version of this work appeared on arXiv.} 
that $M_2(|\tilde\psi\rangle)-4\le M_2^{\rm NL}$ for every pure state~\cite{Sierant:2026lvg}. This gives:}
\CHANGE{\begin{corollary}\label{cor:bracket}
With $|\tilde\psi\rangle$ the computational-basis representative of the Schmidt spectrum of $|\psi\rangle$,
\begin{equation}
\begin{split}
 &\max\bigl\{-\log_2(1-4F),\,M_2(|\tilde\psi\rangle)-4\bigr\}\\
 &\qquad\le M_2^{\rm NL}\le\min\bigl\{M_2(|\tilde\psi\rangle),\,M_2\bigr\}.
\end{split}
 \label{eq:bracket}
\end{equation}
\end{corollary}}
\CHANGE{The two lower bounds in Eq.~\eqref{eq:bracket} are complementary. Since $F<27/256$, the anti-flatness bound never exceeds $\log_2(64/37)\approx0.79$ bits\footnote{\CHANGE{Let $p=\lambda_{\max}$ and $S_n=\mathrm{Tr}\,\rho_A^n$. Since $\lambda_i\le p$, $S_3\le pS_2$, so $F=S_3-S_2^2\le f(S_2)$ with $f(x)=px-x^2=p^2/4-(x-p/2)^2$; moreover $S_2\ge p^2$. For $p\ge\frac12$: $S_2-\frac p2\ge p^2-\frac p2=p(p-\frac12)\ge0$, so $(S_2-\frac p2)^2\ge(p^2-\frac p2)^2$ and $F\le p^2/4-(p^2-\frac p2)^2=p^3-p^4\equiv g(p)$; $g'(p)=p^2(3-4p)=0$ at $p=\frac34$, $g''(\frac34)=-\frac94<0$, and $g(\frac34)=\frac{27}{256}$ exceeds $g(\frac12)=\frac1{16}$ and $g(1)=0$. For $p<\frac12$: $F\le f(S_2)\le p^2/4<\frac1{16}$. Hence $F\le\frac{27}{256}$ and $1-4F\ge\frac{37}{64}$.}}, and it decays at least as fast as $2^{-M_2^{\rm NL}/2}$ along any family in which the non-local magic diverges, since $F\le\lambda_{\max}^2\le4^{-D_{\min}^{\rm NL}}\le2^{-M_2^{\rm NL}/2}$, where $D_{\min}^{\rm NL}$ is the non-local min-relative entropy of non-stabilizerness, whose closed form~\cite{Sierant:2026qji} gives $\lambda_{\max}\le2^{-D_{\min}^{\rm NL}}$, and $M_2\le4D_{\min}$. In that regime, the bound of Ref.~\cite{Sierant:2026lvg} is more accurate. Conversely, $M_2(|\tilde\psi\rangle)-4$ is redundant whenever $M_2(|\tilde\psi\rangle)<4$, which holds for all lattices and couplings studied here, so that $-\log_2(1-4F)$ is the only non-trivial lower bound available for our data.}

We evaluate Eq.~\eqref{eq:anti-flatness-def} at the symmetric vertical cut through the center of the
lattice. The Schmidt decomposition of the converged MPS gives the eigenvalues of
$\hat\rho_A$ directly, $\lambda_i = s_i^2$ with $s_i$ the Schmidt coefficients,
so that $\operatorname{Tr}\hat\rho_A^n = \sum_i \lambda_i^n$ and the
lower bound $-\log_2(1-4F)$ follows in a straightforward manner.

\paragraph{Encoding-invariance and preference of non-local quantities.}
The three measures respond very differently to the way the gauge theory is
mapped onto qubits, and this is central to our conclusions. Each gauge-invariant
dressed site has a physical dimension $30$, $13$, or $6$ (bulk, edge, corner),
which can be embedded into $5$, $4$, or $3$ qubits [Eq.~\eqref{eq:dims}]; this embedding
is not unique. Because SRE is invariant under Clifford operations but \emph{not}
under general unitaries, the full magic $M_2$ depends on this choice: a different
site-local encoding would generally yield a different $M_2$ for the \emph{same} physical state. By contrast, the entangling cut lies between sites, so every re-encoding is a product of unitaries supported entirely within $A$ or within $B$. The non-local magic $M_2^{\mathrm{NL}}$, defined through the minimization over $U_A\otimes U_B$ in Eq.~\eqref{eq:MNL}, is therefore invariant under any such re-encoding, and the anti-flatness $F$, depending only on the Schmidt spectrum, is similarly invariant. 

In the dressed-site approach, absorbing the rishons into the sites renders every hopping and plaquette operator
even under the on-site fermion parity, so no Jordan--Wigner strings connect
distinct sites and the vertical cut is a genuine tensor-product bipartition
$\mathcal H=\mathcal H_A\otimes\mathcal H_B$. A site-local re-encoding is thus a
true local unitary $U_A\otimes U_B$, and the invariance of $M_2^{\mathrm{NL}}$ and
$F$ follows without any string ambiguity.

Hence, $M_2^{\mathrm{NL}}$ and its anti-flatness dependent bound\CHANGE{, as well as our estimate $M_2(|\tilde\psi\rangle)$, which depends only on the Schmidt spectrum,} are intrinsic to the state and the bipartition, whereas $M_2$ carries a residual dependence on the qubit encoding convention. This gives another reason, beyond the numerical convergence discussed in Sec.~\ref{sec:results} below, to prioritize non-local quantities as the physical diagnostics of the gauge–matter delocalization crossover.

\paragraph{Entanglement entropy and edge modes.}
Entanglement entropy for gauge theories has a long history and is subtle~\cite{Sorkin:1984kjy, Balachandran:1994vi,Velytsky:2008rs,Buividovich:2008gq, Donnelly:2011hn, Casini:2013rba,Gromov:2014kia,Donnelly:2014gva,Aoki:2015bsa,Soni:2015yga,Itou:2015cyu}. 
On a spatial lattice, the gauge degrees of freedom live on links, and the physical space is the gauge-invariant subspace annihilated by the Gauss law at every vertex, $\mathcal H_{\rm phys}=\{|\psi\rangle:G_x|\psi\rangle=0\ \forall\,x\}$.
Partitioning links into regions $A$ and $B$, the entangling surface runs through
the boundary vertices and $\mathcal H_{\rm phys}$ does \emph{not} factorize: $G_x$
at a boundary vertex mixes links of both regions, so no gauge-invariant state
splits as $\mathcal{H} = \mathcal{H}_{A}\otimes\mathcal{H}_{B}$. The extended-space
prescription instead embeds $\mathcal H_{\rm phys}$ into
$\mathcal H_{\rm ext}=\mathcal H_A^{\rm ext}\otimes\mathcal H_B^{\rm ext}$ by
dropping the Gauss law on the cut, so that each region carries a product over its
own links; one then sets
$\rho_A=\mathrm{Tr}_{\mathcal H_B^{\rm ext}}|\psi\rangle\langle\psi|$ and
$S=-\mathrm{Tr}\,\rho_A\log_{2}\rho_A$. The price of factorization is the
admission of gauge-non-invariant boundary states---the \emph{edge modes}---and the
entanglement entropy decomposes into three physically distinct
contributions (referred to as classical, quantum, and edge)~\cite{Donnelly:2011hn},
\begin{equation}
    S = \underbrace{-\sum_j p_j \log_2 p_j}_{\text{classical}} + \underbrace{\sum_j p_j\, S_j}_{\text{quantum}} + \underbrace{\sum_j p_j \log_2 d_j}_{\text{edge}},
    \label{eq:EE_decomposition}
\end{equation}
where $j\in\{0,\tfrac12\}$ labels the boundary irreducible representations populated at the cut, $p_j$
is the probability of irreducible representation $j$, $S_j$ is the entanglement entropy within the
$j$-sector, and $d_j$ is its dimension ($d_0=1$, $d_{1/2}=2$), so the third term (edge term) reduces to $p_{1/2}\log_2 2 = p_{1/2}$ for the hardcore gluon truncation. 

In the dressed-site approach, the situation simplifies. The vertex Gauss law is
\emph{already solved} into a locally gauge-invariant basis of a fixed dimension, so
there is no residual vertex obstruction; the only remaining constraint is the
link constraint (or link law) since a physical link is two rishons, one per end, locked by matching
flux [Eq.~\eqref{eq:linkconstraint}]. On a dressed-site lattice the natural cut is
a set of bonds, and these link constraints are the entire content of
$\mathcal H\neq\mathcal H_A\otimes\mathcal H_B$. Extending the space here means
splitting each cut link into its two rishons and dropping the matching
constraint; the unlocked rishons are the edge modes.

Lastly, to make a connection between the measures of magic and the physics of the model, we focus on the standard gauge-invariant observables in the truncated lattice gauge theory.
The gauge-field observables we compute are the electric
energy density, the normalized magnetic plaquette, and the magnetic energy defined as:
\begin{align}
  \langle E^2\rangle
  &= \frac{1}{N_{\ell}}\sum_{\ell}\big\langle \hat E_{\ell}^{2}\big\rangle,
  \label{eq:E2}\\[2pt]
  \Ree\langle B_p\rangle
  &= \CHANGE{\frac{1}{N_{\square}}}\sum_{\square}
     \Ree\big\langle \Tr\Box\big\rangle,
  \label{eq:Bp}\\[2pt]
  \langle B^2\rangle
  &= \frac{1}{2N_{\square}}\sum_{\square}
     \Big(1-\CHANGE{\Ree\big\langle \Tr\Box\big\rangle}\Big) \nonumber \\
   & = \tfrac12\big(1-\Ree\langle B_p\rangle\big),
  \label{eq:B2}
\end{align}
where $\Box=U_{\bj,\bj+\mux}\,U_{\bj+\mux,\bj+\mux+\muy}\,
U^{\dagger}_{\bj+\muy,\bj+\mux+\muy}\,U^{\dagger}_{\bj,\bj+\muy}$ is the plaquette
of Eq.~\eqref{eq:Hpuredim}, $\hat E_{\ell}^{2}$ is the color-electric Casimir on
link $\ell$, and $N_{\ell}$, $N_{\square}$ are the numbers of links and plaquettes (with open
boundaries on an $L\times L$ lattice, \CHANGE{$N_{\ell}=2L(L-1)$ and} $N_{\square}=(L-1)^2$).
The electric term of Eq.~\eqref{eq:Hdim_main} favors $\langle E^2\rangle=0$ at
strong coupling, whereas the magnetic term lowers its energy by ordering the
plaquettes ($\Ree\langle B_p\rangle\to1$) at weak coupling; the competition
between the two produces the crossover that we find in our results in Sec.~\ref{sec:results}. 

\section{\label{sec:results}Results}

We use a one-dimensional tensor train built as the matrix product state (MPS) of the two-dimensional lattice in the column-major ordering, as shown in Fig.~\ref{fig:dressed_site}. In a previous study of dressed-site $\SUtwo$ in the hardcore gluon approximation~\cite{Cataldi:2023xki}, tree tensor networks (TTN) were used. Our choice of MPS is partially motivated by recent work~\cite{Barthel:2026ome}, which showed that the scaling of MPS is superior to TTN for two-dimensional quantum systems with area-law entanglement (gapped systems) on open square lattices. We evaluate the full SRE by \emph{perfect} (direct) sampling of the Pauli strings from the matrix product  state~\cite{Lami:2023naw} inspired by an earlier sampling procedure~\cite{Ferris:2012pbh} exploiting the MPS structure. 
For bigger lattice sizes, we only compute \CHANGE{the non-local quantities $M_2(|\tilde\psi\rangle)$ and $-\log_2(1-4F)$} since the full SRE becomes computationally expensive. The lack of access to full magic does not alter the overall conclusions. As discussed earlier, non-local magic is the non-stabilizerness that remains once all single-site contributions have been optimized away, so it measures only the magic stored in inter-site correlations across a partition. Any local product unitary, an encoding change in particular, acts within a single site and therefore cannot change non-local magic, leading to its basis- and encoding-independent nature.

\CHANGE{Gauss's law is solved exactly on each dressed site, the link constraint of Eq.~\eqref{eq:linkconstraint} is imposed by an energy penalty that commutes with the Hamiltonian, and the baryon number is imposed exactly as a $U(1)$ charge of the MPS. Ground states are obtained with two-site DMRG (with the density-matrix mixer, at most $28$ sweeps, energy tolerance $10^{-12}$), continuing upward in $g^2$ from the converged state at the previous coupling. We have checked the dressed-site Hamiltonian against an independent construction in the original link basis by comparing full spectra on the $2\times2$ lattice.}

\begin{figure*}
    \centering
    \includegraphics[width=0.98\linewidth]{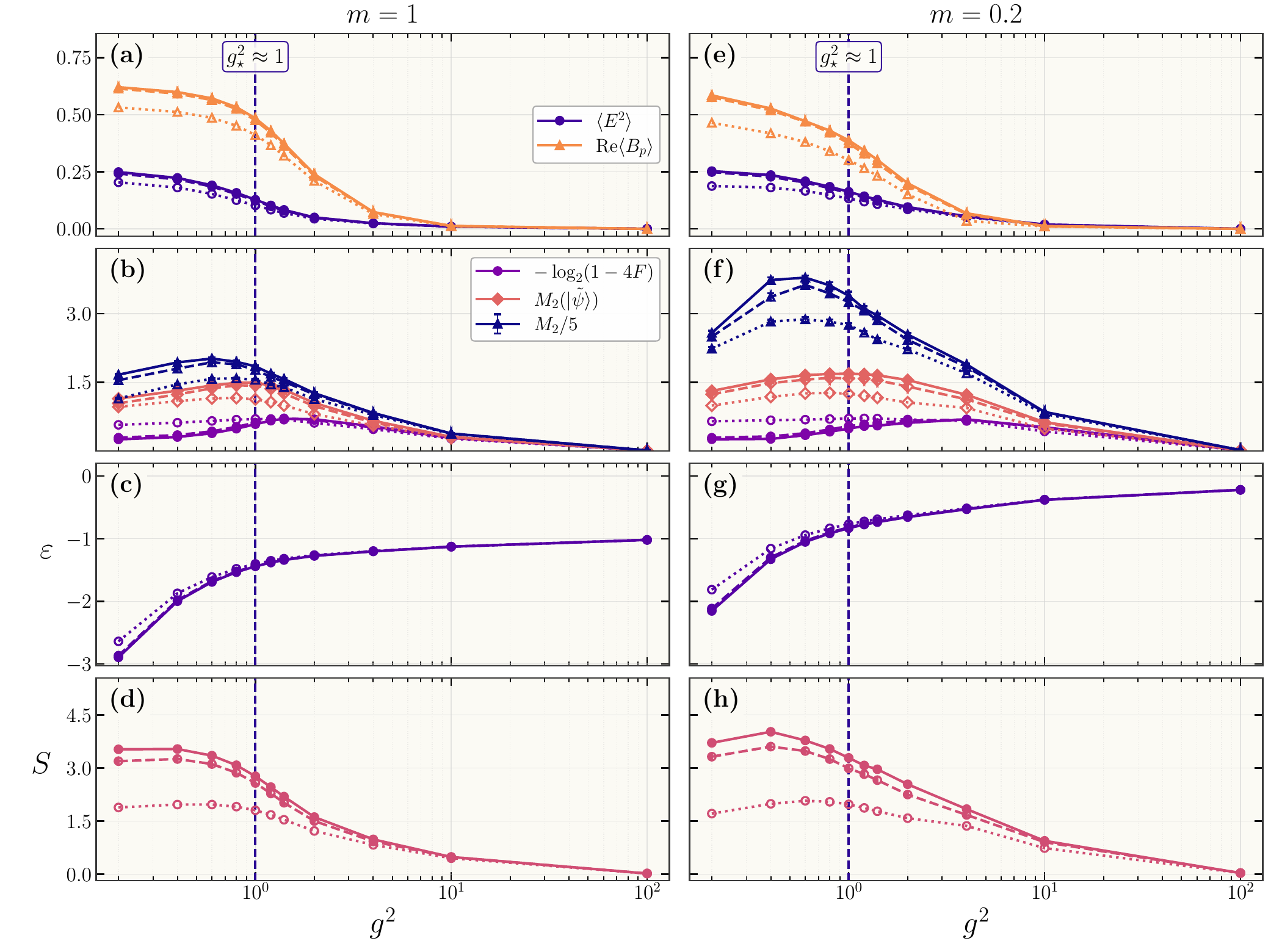}
    \caption{Coupling dependence and bond-dimension convergence on the $4\times4$ lattice with $\chi=32$ ($\Circle$, \CHANGE{dotted}), $64$ ($\LEFTcircle$, \CHANGE{dashed}), and $96$ ($\CIRCLE$, solid lines) \CHANGE{for $m=1$ [left column, (a)--(d)] and $m=0.2$ [right column, (e)--(h)]. Panels in the same row share the vertical axis.} The vertical dashed line marks the gauge–matter delocalization crossover at $g_\star^2 \sim 1$. \CHANGE{(a,e)}~Gauge observables: $\langle E^2\rangle$ and $\Ree\langle B_p\rangle$ both increase (while $\langle B^2\rangle$ decreases as per Eq.~\eqref{eq:B2}) as we approach $g^2 \to 0$. The plaquette-based observables carry the strongest $\chi$-dependence. \CHANGE{(b,f)}~Non-stabilizerness measures \CHANGE{$M_2/5$ (dark blue triangles), $M_2(|\tilde\psi\rangle)$ (salmon diamonds), the conjecture-free upper bound on $M_2^{\rm NL}$, and the lower bound $\max\{-\log_2(1-4F),\,M_2(|\tilde\psi\rangle)-4\}$ of Eq.~\eqref{eq:bracket} (purple circles), which equals $-\log_{2}(1-4F)$ at every point shown since $M_2(|\tilde\psi\rangle)<4$. The salmon and purple curves are the upper and lower ends of Eq.~\eqref{eq:bracket}; the exact $M_2^{\rm NL}$ lies between them and equals the salmon curve under the conjecture of Ref.~\cite{Liu:2026nah}. We have scaled down the full SRE $M_2$ by a factor of 5 for improved visibility.} \CHANGE{(c,g)}~Ground state energy density $\varepsilon$. \CHANGE{(d,h)}~Gauge-invariant entanglement entropy $S$ across the central cut.
    }
    \label{fig:4x4_Dcheck}
\end{figure*}

\begin{figure*}
    \centering
    \includegraphics[width=0.98\linewidth]{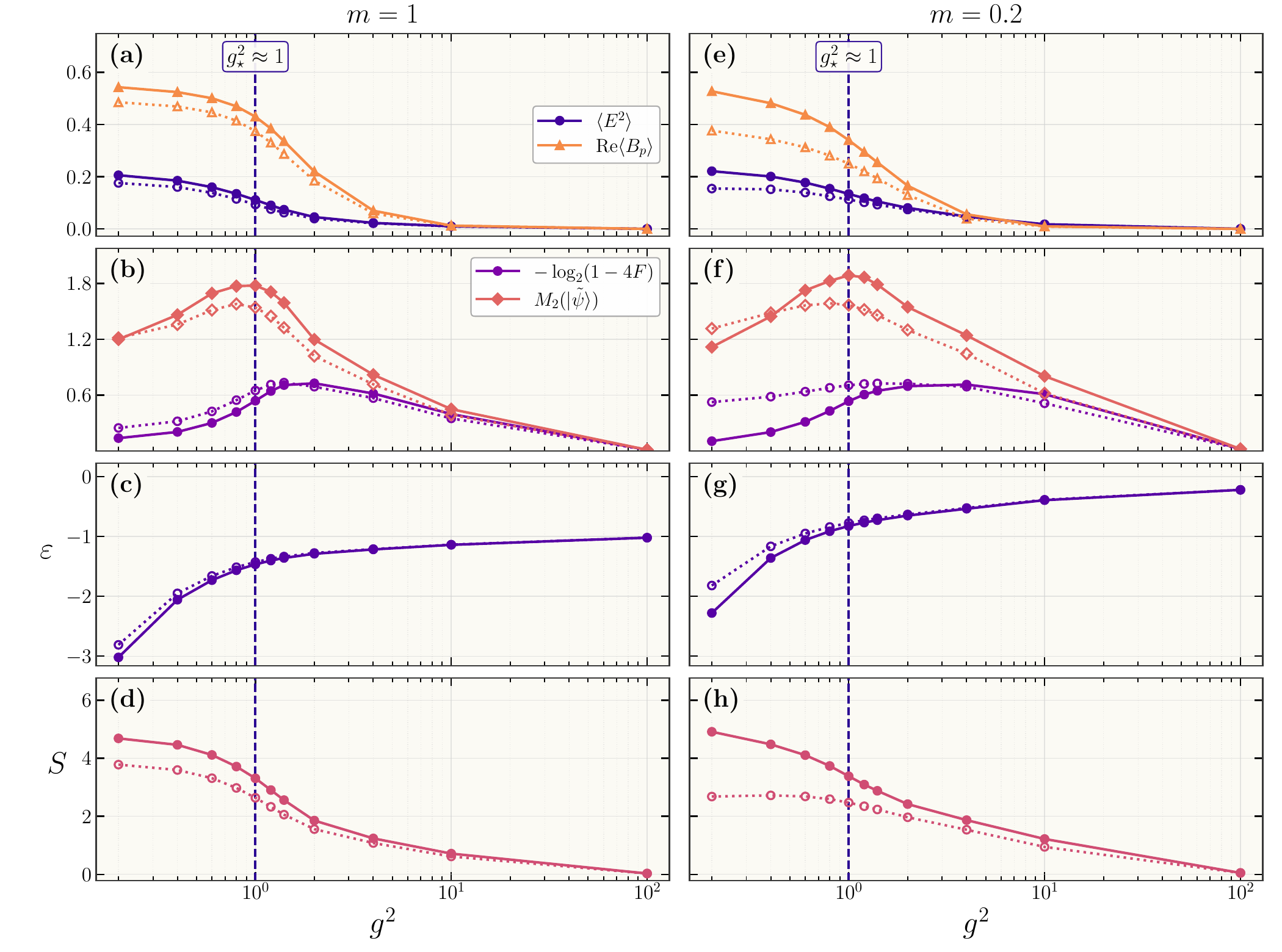}
    \caption{Coupling dependence and bond-dimension convergence on the $6\times6$ lattice with $\chi=64$ ($\Circle$, \CHANGE{dotted}), and $\chi=128$ ($\CIRCLE$, solid lines) versus $g^2$ \CHANGE{for $m=1$ [left column, (a)--(d)] and $m=0.2$ [right column, (e)--(h)]. Panels in the same row share the vertical axis.} The vertical dashed line marks the gauge–matter delocalization crossover at $g_\star^2 \sim 1$. \CHANGE{(a,e)}~Gauge observables: $\langle E^2\rangle$ and $\Ree\langle B_p\rangle$ both increase as we approach $g^2 \to 0$. \CHANGE{(b,f)}~Non-stabilizerness measures \CHANGE{$M_2(|\tilde\psi\rangle)$, the conjecture-free upper bound on $M_2^{\rm NL}$, and the lower bound of Eq.~\eqref{eq:bracket}, which equals $-\log_{2}(1-4F)$ at every point shown; the exact $M_2^{\rm NL}$ lies between the two curves}. \CHANGE{(c,g)}~Ground state energy density $\varepsilon$. \CHANGE{(d,h)}~Gauge-invariant entanglement entropy $S$ across the central cut. }
    \label{fig:main_results_6x6}
\end{figure*}

\begin{figure*}
    \centering
    \includegraphics[width=0.98\linewidth]{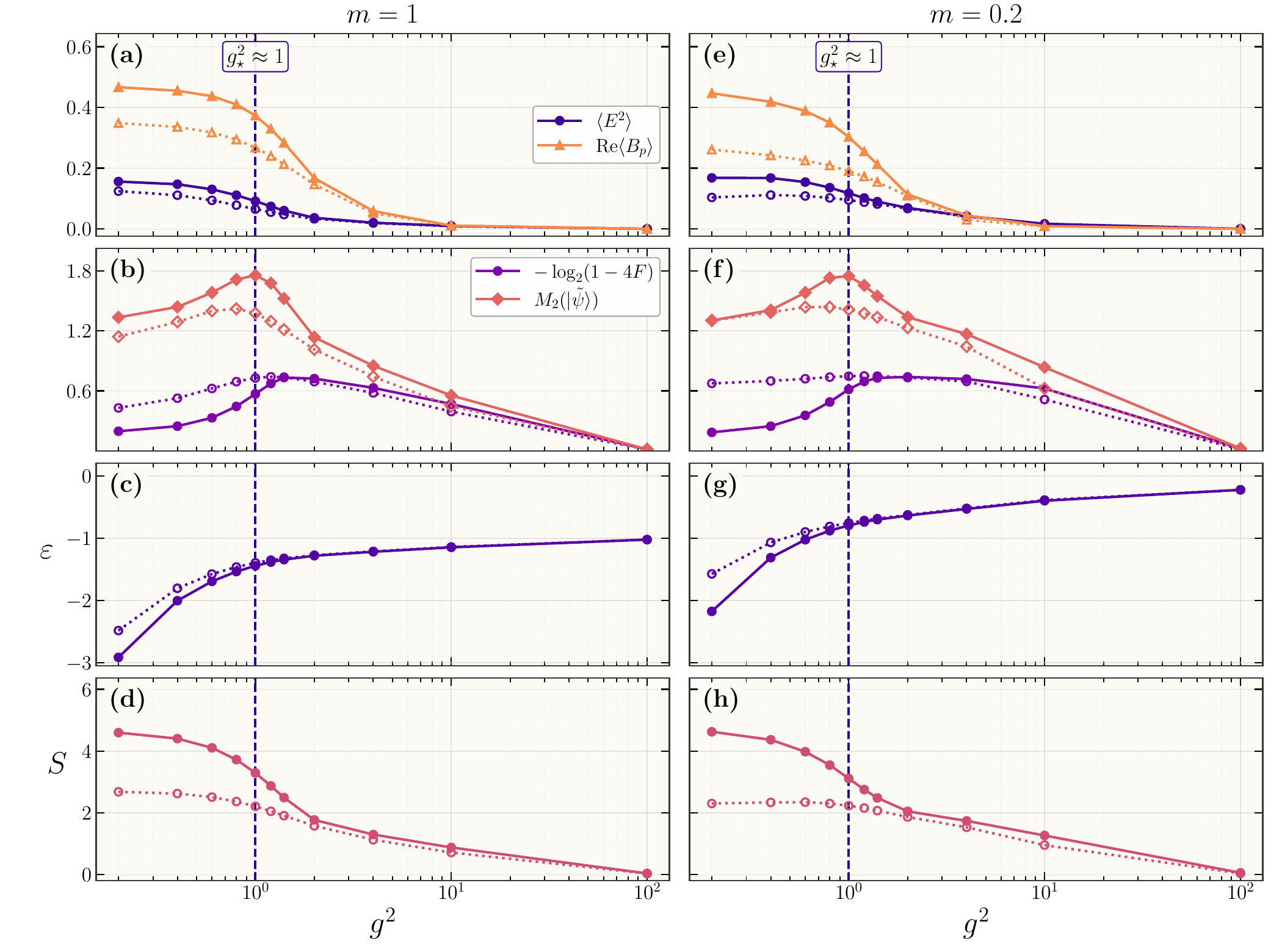}
    \caption{\CHANGE{Same as Fig.~\ref{fig:main_results_6x6} for the $8\times8$ lattice with $\chi=64$ ($\Circle$, dotted) and $\chi=128$ ($\CIRCLE$, solid lines), for $m=1$ [left column, (a)--(d)] and $m=0.2$ [right column, (e)--(h)]. For $\chi=128$, the lower-energy state of two runs is shown at each $g^2$ (see text).}}
    \label{fig:main_results_8x8}
\end{figure*}

\CHANGE{We consider two values of the bare mass, $m=1$ and $m=0.2$, shown in the left and right columns of Figs.~\ref{fig:4x4_Dcheck}, \ref{fig:main_results_6x6}, and \ref{fig:main_results_8x8}, respectively; unless stated otherwise, the discussion refers to $m=1$.} We start by investigating the dependence on bond dimension on the smallest lattice ($4 \times 4$) considered in this work.  The results are shown in Fig.~\ref{fig:4x4_Dcheck}. The gauge observables associated with the plaquette (magnetic term) show the strongest dependence on the bond dimension\CHANGE{, while all observables vary smoothly with $g^2$ for every $\chi$.}

\begin{figure*}
    \centering
    \includegraphics[width=0.98\linewidth]{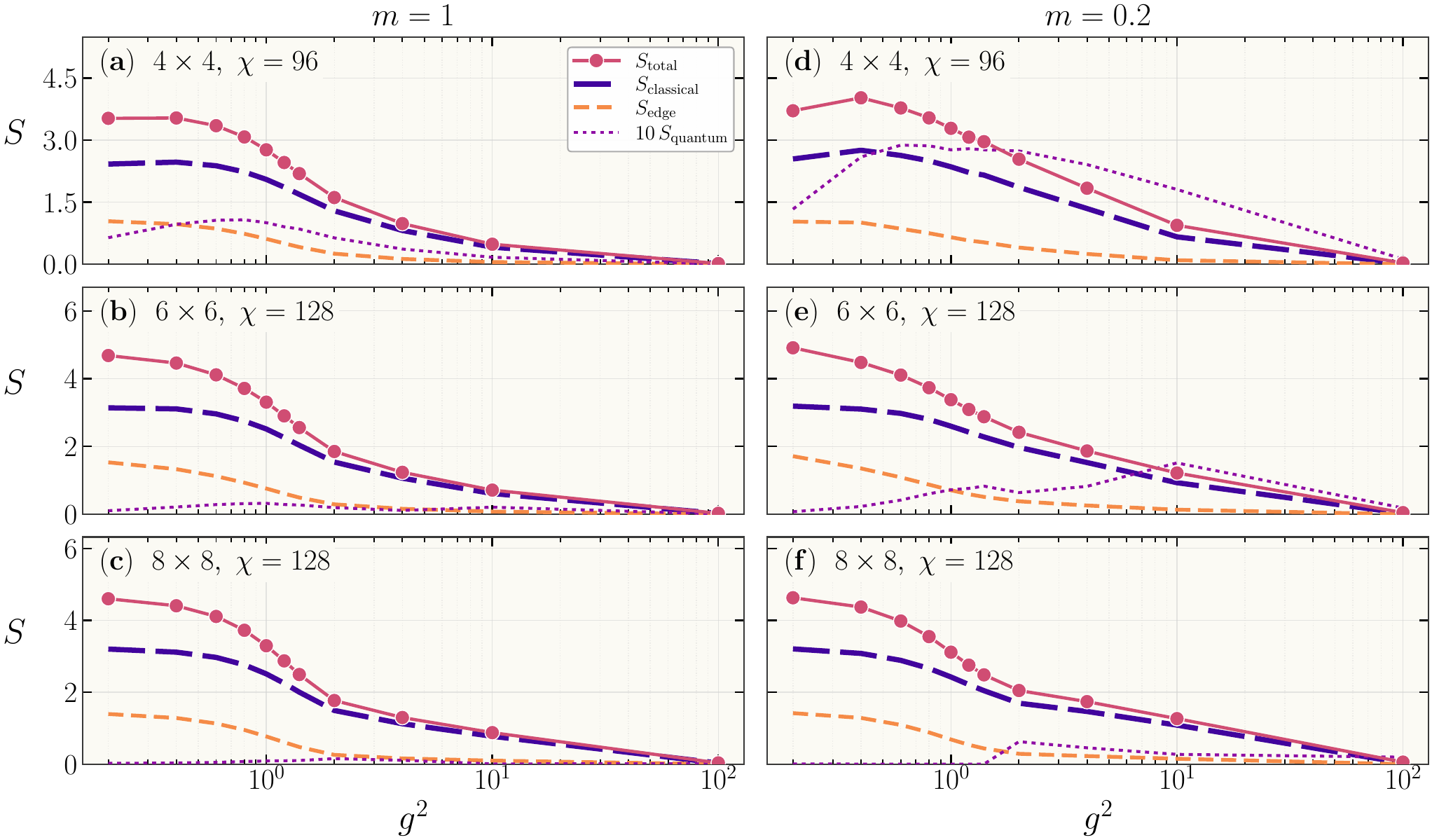}
    \caption{The individual contributions to the total entanglement entropy defined in Eq.~\eqref{eq:EE_decomposition} with largest $\chi$ \CHANGE{for the $4\times4$ ($\chi=96$), $6\times6$, and $8\times8$ ($\chi=128$) lattices (top to bottom), for $m=m_0a=1$ [left column, (a)--(c)] and $m=0.2$ [right column, (d)--(f)]. Panels in the same row share the vertical axis.} The quantum part of the entropy has been scaled by 10$\times$ for improved visibility.}
    \label{fig:ee_decomposition_both_lattices}
\end{figure*}

Increasing $\chi$ from $64$ to $96$ results in only small changes in the magic observables and in the entanglement\CHANGE{: in the crossover region $0.6\le g^2\le1.2$, $M_2$ and $M_2(|\tilde\psi\rangle)$ change by $3$--$5\%$ and $S$ by $7$--$8\%$}.
\CHANGE{The most interesting aspect of the results on the smaller lattice is that $M_2(|\tilde\psi\rangle)$ has a broad maximum at $0.8\lesssim g^2\lesssim1$ for all three bond dimensions; for $\chi=64$ and $\chi=96$, its values at $g^2=0.8$ and $g^2=1$ agree to within $0.5\%$, and the location of the maximum does not move between the two, signaling convergence. By contrast, the full SRE $M_2$ peaks at $g^2\approx0.6$ for $\chi=64$ and $96$, the lower bound of Eq.~\eqref{eq:bracket} peaks at $g^2\approx1.4$, and the entanglement entropy shows no peak near the crossover; its steepest decrease occurs between $g^2=1$ and $g^2=1.4$.}
In our Hamiltonian convention, the crossover discussed in Refs.~\cite{Cataldi:2023xki, Rouxinol:2026vjl} should occur at $g^2 \sim 1$; this is picked up earlier (in terms of bond dimension scaling) and more accurately by \CHANGE{$M_2(|\tilde\psi\rangle)$} than any other observable. 
\CHANGE{For all lattices, masses, and couplings considered in this work, $M_2(|\tilde\psi\rangle)\le1.9<4$, so the lower bound $\max\{-\log_2(1-4F),\,M_2(|\tilde\psi\rangle)-4\}$ of Eq.~\eqref{eq:bracket} is attained by its anti-flatness term at every data point; this is the curve shown in panels (b) and (f) of Figs.~\ref{fig:4x4_Dcheck}, \ref{fig:main_results_6x6}, and \ref{fig:main_results_8x8}. Because the plotted $M_2(|\tilde\psi\rangle)$ is an unconditional upper bound on $M_2^{\rm NL}$, the fact that it lies above this lower bound for all $g^2$ is guaranteed by Eq.~\eqref{eq:bracket}; the exact $M_2^{\rm NL}$ lies between the two curves.} 
\CHANGE{The two ends of the bracket, $M_2(|\tilde\psi\rangle)$ and $-\log_2(1-4F)$, approach each other in the strong-coupling limit $g^2\to\infty$, where the ground state approaches a product (stabilizer) state.}

\CHANGE{For the lighter mass $m=0.2$ [Fig.~\ref{fig:4x4_Dcheck}(e)--(h)], the plaquette expectation value is smaller and the electric energy is larger. The full SRE nearly doubles, reaching $M_2\approx19$ at $g^2=0.6$ for $\chi=96$, compared with $M_2\approx10$ for $m=1$. $M_2(|\tilde\psi\rangle)$ increases much less, by $11$--$13\%$ at its maximum, and its broad maximum remains at $0.8\lesssim g^2\lesssim1.2$. The full SRE, which also counts magic that can be removed by on-site unitaries, is therefore far more sensitive to the matter sector than $M_2(|\tilde\psi\rangle)$, whose maximum continues to track the crossover. The lower bound has a broad maximum at larger couplings, $2\lesssim g^2\lesssim4$ for $\chi=64$.}

As we increase the lattice size to $6 \times 6$, the entanglement grows and $\chi=96$ is insufficient for the proper convergence of the ground state. Hence, we use a maximum bond dimension of $\chi=128$. The decomposition of the total gauge-invariant entanglement entropy shown in Fig.~\ref{fig:ee_decomposition_both_lattices} further clarifies this growth. The edge contribution increases steadily from $4\times 4$ to $6\times 6$ due to more bonds being severed. \CHANGE{On $8\times8$, the edge contribution at $\chi=128$ ($\approx1.4$) falls below its $6\times6$ value, an indication that $\chi=128$ does not converge the entanglement on the largest lattice (see below)}.
We also find that the quantum component of the entanglement entropy has stronger bond-dimension dependence compared to the other two counterparts, but is small compared to other contributions\CHANGE{; it is larger for the lighter mass, by a factor of $3$--$5$ at its maximum}. This part of entropy has been shown to correspond to distillable entropy~\cite{ VanAcoleyen:2015ccp} and it would be interesting to see how this behaves when higher $\jmax$ is used. 

We consider the same set of observables as $4 \times 4$, except the extensive SRE $M_2$ of the ground state. We evaluate the dependence on various observables by comparing it with a smaller bond dimension of $\chi=64$. Most of the observables change considerably between $\chi = 64$ and $\chi=128$. For example, while the peak of \CHANGE{$M_2(|\tilde\psi\rangle)$} is at $g^2 \approx 0.8$ with $\chi=64$, it moves to $g^2\approx1$ for $\chi=128$, consistent with the gauge–matter delocalization crossover identified in Ref.~\cite{Cataldi:2023xki}. In addition, the dependence of the entanglement entropy on $g^2$ is very similar to what was obtained for 1+1-dimensional $\SUtwo$ in Ref.~\cite{Jha:2026ror}, and the steepest gradient in entropy occurs close to $g^2 \sim 1$. The results are shown in Fig.~\ref{fig:main_results_6x6}.
\CHANGE{As for $4\times4$, the curves satisfy Eq.~\eqref{eq:bracket} for the entire range of $g^2$. The maximum of the lower bound, $\approx0.73$, again comes close to the ceiling $\log_2(64/37)\approx0.79$, while $M_2(|\tilde\psi\rangle)\lesssim1.9$ stays well below the $4$ bits at which the bound $M_2(|\tilde\psi\rangle)-4$ of Ref.~\cite{Sierant:2026lvg} becomes non-trivial. For $m=0.2$ [Fig.~\ref{fig:main_results_6x6}(e)--(h)], the maximum of $M_2(|\tilde\psi\rangle)$ again moves from $g^2\approx0.8$ at $\chi=64$ to $g^2\approx1$ at $\chi=128$, with a slightly larger value, so its location is insensitive to the fermion mass.} The location of the crossover, $g_{\star}$, we found using \CHANGE{$M_2(|\tilde\psi\rangle)$} is consistent (numerically close) with results obtained in Ref.~\cite{Rouxinol:2026vjl} with untruncated $\SUtwo$ with a continuous quantum Monte Carlo approach.

\CHANGE{Finally, we consider the $8\times8$ lattice (288 qubits) with $\chi=64$ and $\chi=128$, shown in Fig.~\ref{fig:main_results_8x8}. Here $\chi=128$ is clearly insufficient to converge the ground state at small $g^2$: at $g^2=0.2$ the entanglement entropy grows substantially between $\chi=64$ and $\chi=128$, and at $\chi=128$ it still lies below its $6\times6$ value. In this case, $M_2(|\tilde\psi\rangle)$ at $\chi=128$ has its maximum at $g^2\approx1$ for both masses ($\approx1.76$ for $m=1$ and $\approx1.75$ for $m=0.2$), as on $6\times6$, whereas at $\chi=64$ the maximum lies at $g^2\approx0.8$ ($m=1$) and $0.6\lesssim g^2\lesssim0.8$ ($m=0.2$). The peak values lie slightly below the $6\times6$ ones, again reflecting the truncation. However, the peak remains robust to the change in $\chi$ even though the ground state has likely not converged fully.}

The observation that the same crossover appears in the hardcore ($j_{\max}=1/2$) truncation and in the untruncated theory~\cite{Rouxinol:2026vjl} in the physical regime is an encouraging indication that
the qualitative physics is largely not very sensitive to the representation cutoff, as one would expect if a low-lying truncation already captures the relevant low-energy physics thanks to universality. However, this requires 
demonstrating cutoff-independent scaling as $j_{\max}$ is increased, which is beyond the scope of this work and might be extremely difficult to obtain using current methods.

\section{Summary and discussion}

In this paper, we examine the non-stabilizerness associated with the ground state of $\SUtwo$ gauge theory in $2+1$ dimensions on a square lattice with open boundary conditions working in the hardcore gluon approximation. Using one-dimensional ordering of matrix product states with bond dimension up to $\chi=128$ on lattices up to \CHANGE{$8\times8$ and for two values of the fermion mass}, we computed gauge-invariant observables together with the stabilizer R\'enyi entropy (SRE), its non-local component, and a \CHANGE{lower bound with optimal constant} to non-local magic in terms of anti-flatness. We find that the non-local magic evaluated in the truncated Schmidt basis~\cite{Liu:2026nah}\CHANGE{, $M_2(|\tilde\psi\rangle)$, which bounds $M_2^{\rm NL}$ from above and coincides with it under the conjecture of Ref.~\cite{Liu:2026nah},} is the sharpest probe of the gauge–matter delocalization crossover: it retains a clear signal at substantially smaller bond dimension than the full SRE, the derived lower bound, or the gauge-invariant bipartite entanglement entropy.

The hardcore ($j_{\max}=1/2$) truncation already yields a $30$-dimensional local Hilbert space on bulk sites, yet this remains far from what is required to reach the continuum limit. The central open direction is therefore to go beyond the hardcore limit retaining higher irreducible representations of the dressed-site basis and to study systematically how both the quantum-information and gauge-theoretic observables evolve as the continuum is approached in this $\SUtwo$ Hamiltonian lattice gauge theory. Our preliminary estimates give a local dressed-site dimension of $168$ for $j_{\max}=1$ and $600$ for $j_{\max}=3/2$. While the smallest cutoff needed to reach within a few percent of the continuum is not known \emph{a priori}, earlier work suggests $j_{\max}\ge 2$~\cite{Bazavov:2019qih, Zache:2023dko}. Computing the non-local magic in this regime would connect directly to recent results arguing that the physical states
of target quantum field theories must carry a very large, and possibly divergent, amount of magic~\cite{Benedetti:2026mfy, Moosa:2026nqz}. This is consistent with our earlier conclusion~\cite{Jha:2026ror} that approaching the continuum limit of a Hamiltonian lattice gauge theory drives the dimensionless couplings into a regime where the ground state possesses increasingly large magic.
\CHANGE{In that regime, the anti-flatness bound alone is not informative: it is capped at $\log_2(64/37)\approx0.79$ bits and decays exponentially in $M_2^{\rm NL}$ along any family with divergent non-local magic, whereas the bound $M_2(|\tilde\psi\rangle)-4$ of Ref.~\cite{Sierant:2026lvg} captures the divergence up to $O(1)$. The bracket of Eq.~\eqref{eq:bracket}, which requires only the entanglement spectrum, therefore provides a conjecture-free handle on the non-local magic in both regimes.}

It would therefore be valuable to test whether the same behavior emerges in the $2+1$-dimensional theory studied here as the irreducible representation cutoff is increased and we approach the continuum limit, and ultimately in $3+1$-dimensional gauge theories. Such studies would help clarify the interplay between classical simulability, gauge invariance, and quantum error correction, and the role of non-stabilizerness in representing the physical states associated with the stable particles observed in experiment. More broadly, they reinforce that the intersection of quantum information and quantum field theory continues to expose deep connections between computational complexity and emergent physical structure.

A further natural extension is to move from the ground state magic considered here to the magic of time-evolved states, where some progress has recently been made in $1+1$ dimensions~\cite{Bhakuni:2026yve}. In particular, probing scattering in one and two spatial dimensions~\cite{Jha:2024jan, Pavesic:2025nwm} with tensor networks and tracking the growth of magic as wave packets evolve across both integrable and non-integrable regimes would sharpen our understanding of when classical methods remain reliable and where quantum computation is expected to be 
advantageous.

\section*{Data Availability Statement}

The code and data used in this work built using TeNPy~\cite{Hauschild_2024} is available here: \href{https://github.com/rgjha/2dSU2MagicEE}{\texttt{https://github.com/rgjha/2dSU2MagicEE}}. 

\section*{Acknowledgments}
 
R.G.J., J.I.T., M.A., B.N.B., and A.F.K. were supported by the U.S. Department of Energy, Advanced Scientific Computing Research, under
contract number DE-SC0025384. The authors acknowledge assistance of Anthropic’s Claude Opus 4.8 and Fable 5.0 in code optimizations, assistance related to Motzkin numbers, and refining the text for improved readability. We acknowledge the use of Claude Fable 5.1 and OpenAI's GPT-6 Astra for assistance in establishing a stronger inequality for the non-local magic. The authors have verified and assume complete responsibility for the results presented. We acknowledge the computing resources provided
by North Carolina State University High Performance Computing Services Core Facility (RRID:SCR 022168).

\bibliographystyle{utphys}
\bibliography{refs}

\clearpage
\onecolumngrid
\appendix
\renewcommand\thefigure{S\arabic{figure}}  
\renewcommand\thetable{S\arabic{table}}  
\setcounter{figure}{0}

\section{\label{app:sec1}Hamiltonian formulation in the dressed basis}
We study $\SUtwo$ gauge theory in $2+1$ dimensions with matter on a two-dimensional square lattice, using the Kogut--Susskind Hamiltonian formulation \cite{Kogut:1974ag}. The gauge fields live on the links, while staggered fermions are placed on the lattice sites. In this study, we use \CHANGE{the dressed-site} formulation of the $\SUtwo$ gauge theory where energy-truncated gauge degrees of freedom are expressed as fermionic rishon modes and merged with neighboring matter sites resulting in purely bosonic operators acting on the gauge-invariant Hilbert space. While the dressed-site approach can be applied up to any irreducible representation of the target gauge group, we consider here the lowest $\SUtwo$ irreducible representations of the gauge link, $j=0$ and $j=1/2$. The electric energy of a link in representation $j$ is $g^2 j(j+1)/2$.
Higher spin representations become increasingly costly at strong and intermediate coupling. This makes the truncation particularly natural in this regime. In particular, recovering the continuum theory would require the representation cutoff to increase, $j_{\max}\to\infty$, as $a\to0$.
Let us now express the Hamiltonian in terms of on-site operators. Consider a square lattice $\Lambda$ with sites: 
\begin{equation}
\bj = (j_x,j_y),
\end{equation}
where $\mux = (1,0)$ and $\muy = (0,1)$ are the unit vectors along the $x$ and $y$ directions, respectively. At every site $\bj$, there are two fermionic modes,
\begin{equation}
  \psi_{\bj,\textcolor{red}{r}},\qquad \psi_{\bj,\textcolor{green}{g}},
\label{eq:matter}
\end{equation}
which transform as the fundamental color doublet of $\SUtwo$ and obey
\begin{equation}
  \{\psi_{\bj,\alpha},\psi_{\bk,\beta}^{\dagger}\}
  =\delta_{\bj\bk}\delta_{\alpha\beta},
  \qquad \alpha,\beta\in\{\textcolor{red}{r},\textcolor{green}{g}\}.
\end{equation}
with local matter Fock space at each site: 
\begin{equation}\label{eq:Hmatter}
 \mathcal H_{\mathrm{m}}
 =\operatorname{span}\left\{
 \ket{\Omega},\,
 \psi_{\textcolor{red}{r}}^{\dagger}\ket{\Omega},\,
 \psi_{\textcolor{green}{g}}^{\dagger}\ket{\Omega},\,
 \psi_{\textcolor{green}{g}}^{\dagger}\psi_{\textcolor{red}{r}}^{\dagger}\ket{\Omega}
 \right\}.
\end{equation}
In the ordered basis of Eq.~\eqref{eq:Hmatter} the matter operators and the matter parity are, 
\begin{equation}
    \psi_{\textcolor{red}{r}} =
    \begin{pmatrix}
        0 & 1 & 0 & 0 \\
        0 & 0 & 0 & 0 \\
        0 & 0 & 0 & -1 \\
        0 & 0 & 0 & 0
    \end{pmatrix},
    \qquad
    \psi_{\textcolor{green}{g}} =
    \begin{pmatrix}
        0 & 0 & 1 & 0 \\
        0 & 0 & 0 & 1 \\
        0 & 0 & 0 & 0 \\
        0 & 0 & 0 & 0
    \end{pmatrix},
    \qquad
    P_\psi = \operatorname{diag}(1,-1,-1,1),
\label{eq:Psi_mat}
\end{equation}
and $\hat{N}= \sum_{\alpha}\psi^\dagger _\alpha \psi_\alpha$. Under color $\SUtwo$, the empty and doubly occupied states are singlets, while the one-particle sector is a doublet:
\begin{equation}
  \mathcal H_{\mathrm m}\simeq \bm 1\oplus \bm 2\oplus \bm 1.
\end{equation}
To make the Hamiltonian site-local, the link parallel transporter \CHANGE{is} written as a product of rishon operators, \CHANGE{one} located at each end of the link,
\begin{equation}
    \hat{U}_{\mathbf{j},\mathbf{j}+\boldsymbol{\mu}}^{\alpha\beta} = \hat{\zeta}_{\mathbf{j},+\boldsymbol{\mu}}^{\alpha} \hat{\zeta}_{\mathbf{j}+\boldsymbol{\mu},-\boldsymbol{\mu}}^{\beta\dagger},
\label{eq:rishon decomposition}
\end{equation}
where $\alpha, \beta \in \{\textcolor{red}{r}, \textcolor{green}{g}\}$. On the rishon space ${\rm span}\big\{\ket{0},\ket{\textcolor{red}{r}},\ket{\textcolor{forestgreen}{g}}\big\}$, we \CHANGE{use}
\begin{equation}
    \zeta^{\textcolor{red}{r}} = \CHANGE{2^{-1/4}}
    \begin{pmatrix}
        0 & 1 & 0 \\
        0 & 0 & 0 \\
        1 & 0 & 0 
    \end{pmatrix},
    \qquad
    \zeta^{\textcolor{green}{g}} = \CHANGE{2^{-1/4}}
    \begin{pmatrix}
        0 & 0 & 1 \\
        -1 & 0 & 0 \\
        0 & 0 & 0 
    \end{pmatrix},
    \qquad
    P = \operatorname{diag}(1,-1,-1),
\label{eq:rishon_mat}
\end{equation}
with generators $\hat{T}^a$ acting as $\frac{1}{2}\sigma^a$ on the
$\{|r\rangle, |g\rangle\}$ block and annihilating $|0\rangle$, so that
\begin{equation}
    \hat{T}^2 = \operatorname{diag}\left(0,\frac{3}{4},\frac{3}{4}\right),
\end{equation}
the Casimir $j(j+1)$ for $j=0,\frac{1}{2}$. \CHANGE{Each $\zeta^{\alpha}$ is odd under $P$ and transforms as a color doublet, $[\zeta^{\alpha},\hat T^{b}]=\frac12\sigma^{b}_{\alpha\alpha'}\zeta^{\alpha'}$; it combines the two rishon species of Ref.~\cite{Cataldi:2023xki} into a single operator, $\zeta^{\alpha}=\zeta^{\alpha}_{A}+\zeta^{\alpha}_{B}$ with $\zeta^{\alpha}_{B}=2^{-1/4}\ket{0}\bra{\alpha}$ and $\zeta^{\alpha}_{A}=i\sigma^{y}_{\alpha\beta}\zeta^{\beta\dagger}_{B}$. The two rishons of a link are tied together by the flux-matching (link) constraint $\hat T^{2}_{\mathbf j,+\bm\mu}=\hat T^{2}_{\mathbf j+\bm\mu,-\bm\mu}$ of Eq.~\eqref{eq:linkconstraint}. On the five-dimensional flux-matched link space ${\rm span}\{\ket{0,0},\ket{\alpha,\beta}\}$ (first entry: rishon at $\mathbf j$, second entry: rishon at $\mathbf j+\bm\mu$), Eq.~\eqref{eq:rishon decomposition} gives
\begin{equation}
  \hat U^{\alpha\beta}
  =\frac{1}{\sqrt2}\Big[\sum_{\alpha'}\varepsilon_{\alpha\alpha'}\,\ket{\alpha',\beta}\bra{0,0}
   +\sum_{\beta'}\varepsilon_{\beta'\beta}\,\ket{0,0}\bra{\alpha,\beta'}\Big],
  \qquad \varepsilon_{rg}=-\varepsilon_{gr}=1,
\label{eq:Umatrix}
\end{equation}
and $\hat U^{\alpha\beta}$ maps the flux-matched space into itself. The factor $2^{-1/4}$ in Eq.~\eqref{eq:rishon_mat} is fixed by requiring the matrix elements of $\hat U$ to be $\pm1/\sqrt2$, the value dictated by the Peter--Weyl normalization of the untruncated transporter, $\langle \tfrac12,m,m'|\hat U^{\alpha\beta}|0,0,0\rangle=\delta_{m\alpha}\delta_{m'\beta}/\sqrt2$; it guarantees $\sum_{\beta}\|\hat U^{\alpha\beta}\ket{0,0}\|^{2}=1$. Equation~\eqref{eq:Umatrix} coincides with the truncated transporter of Ref.~\cite{Cataldi:2023xki}, Eq.~(B33) there.}\footnote{\CHANGE{The rishon matrices quoted in Eq.~(B35) of Ref.~\cite{Cataldi:2023xki} carry a factor $1/\sqrt2$ each; inserted into Eqs.~(B14)--(B15) of that reference they would give matrix elements $1/2$ rather than the $1/\sqrt2$ of its Eq.~(B33). We follow Eq.~(B33). We also note that the magnetic term of Ref.~\cite{Cataldi:2023xki} carries the coefficient $8/g^{2}$ in place of the $1/g^{2}$ of Eq.~(\ref{eq:Hdim_main}), so that values of $g^{2}$ are not directly comparable between the two works.}} By substituting Eq.~\eqref{eq:rishon decomposition} into the hopping term of the Hamiltonian and grouping operators site by site, the interaction simplifies into a product of local operators:

\begin{equation}
  \psi^{\dagger}_{\mathbf j,\alpha}\,
  \hat U^{\alpha\beta}_{\mathbf j,\mathbf j+\bm\mu}\,
  \psi_{\mathbf j+\bm\mu,\beta}
  =\underbrace{\Big(\psi^{\dagger}_{\mathbf j,\alpha}\,
     \hat\zeta^{\alpha}_{\mathbf j,+\bm\mu}\Big)}_{\text{at }\mathbf j}
   \underbrace{\Big(\hat\zeta^{\beta\dagger}_{\mathbf j+\bm\mu,-\bm\mu}\,
     \psi_{\mathbf j+\bm\mu,\beta}\Big)}_{\text{at }\mathbf j+\bm\mu}
  \;=\;\hat Q^{\dagger}_{+\bm\mu}(\mathbf j)\,\hat Q_{-\bm\mu}(\mathbf j+\bm\mu),
\end{equation}
which contracts each color index locally within a single site rather than across the link where,
\begin{equation}
  \hat Q^{\dagger}_{\ell}=\sum_{c=r,g}\big(\psi^{\dagger}_{c}P_\psi\big)
    \,P\cdots P\,\big(\hat\zeta_{c}\big)_{\ell}\CHANGE{,
    \qquad \hat Q_{\ell}=\big(\hat Q^{\dagger}_{\ell}\big)^{\dagger}.}
\label{eq:Qdef}
\end{equation}
Here, $P$ and $P_\psi$ denote the parity operators defined in Eqs.~\eqref{eq:Psi_mat}, \eqref{eq:rishon_mat}, and $\ell$ is the canonical on-site ordering \CHANGE{of the tensor-product slots, $[\text{matter},-x,-y,+x,+y]$}. This construction plays the role of an intra-site Jordan--Wigner string. Although the physical matter and rishon modes are fermionic and should therefore anti-commute, operators acting on different tensor product slots commute by construction. The parity operators provide the necessary sign corrections to recover the proper fermionic anti-commutation relations.  Importantly, $\hat Q^{\dagger}_{\ell}$ contains an even number of fermionic operators (one matter and one rishon operator), so it is even under the total fermion parity. As a result, no Jordan--Wigner strings extending between different sites are required.

\CHANGE{The same site-wise regrouping applies to the Wilson loop,
\begin{equation}
  \Tr\square_{\mathbf j}
  =\sum_{\alpha\beta\gamma\delta}
   \hat U^{\alpha\beta}_{\mathbf j,\mathbf j+\mux}\,
   \hat U^{\beta\gamma}_{\mathbf j+\mux,\mathbf j+\mux+\muy}\,
   \big(\hat U^{\dagger}\big)^{\gamma\delta}_{\mathbf j+\muy,\mathbf j+\mux+\muy}\,
   \big(\hat U^{\dagger}\big)^{\delta\alpha}_{\mathbf j,\mathbf j+\muy},
\end{equation}
with $(\hat U^{\dagger})^{\gamma\delta}=(\hat U^{\delta\gamma})^{\dagger}$. Inserting Eq.~\eqref{eq:rishon decomposition} produces the string of eight rishon operators
\begin{equation}
  \hat\zeta^{\alpha}_{\mathbf j,+x}\,
  \hat\zeta^{\beta\dagger}_{\mathbf j+\mux,-x}\,
  \hat\zeta^{\beta}_{\mathbf j+\mux,+y}\,
  \hat\zeta^{\gamma\dagger}_{\mathbf j+\mux+\muy,-y}\,
  \hat\zeta^{\gamma}_{\mathbf j+\mux+\muy,-x}\,
  \hat\zeta^{\delta\dagger}_{\mathbf j+\muy,+x}\,
  \hat\zeta^{\delta}_{\mathbf j+\muy,-y}\,
  \hat\zeta^{\alpha\dagger}_{\mathbf j,+y}\,.
\end{equation}
Moving the last operator to the second position past six fermionic operators produces no sign, after which the string is a product of four bilinears, one per corner; each bilinear is bosonic, so the corners commute and their order is immaterial:
\begin{align}
  \Tr\square_{\mathbf j}
  &=\hat C_{+x,+y}(\mathbf j)\,
   \tilde C_{-x,+y}(\mathbf j{+}\mux)\,
   \tilde C_{+x,-y}(\mathbf j{+}\muy)\,
   \tilde C_{-y,-x}(\mathbf j{+}\mux{+}\muy),
\label{eq:plaq}\\
  \hat C_{\mu_1\mu_2}&=\sum_{c}\hat\zeta_{c}(\mu_1)\,\hat\zeta^{\dagger}_{c}(\mu_2),\qquad
  \tilde C_{\mu_1\mu_2}=\sum_{c}\hat\zeta^{\dagger}_{c}(\mu_1)\,\hat\zeta_{c}(\mu_2),
\label{eq:cornerdef}
\end{align}
where the products inside $\hat C$ and $\tilde C$ are ordered fermionic products on one site. Note that three of the four corners carry $\zeta^{\dagger}\zeta$ rather than $\zeta\zeta^{\dagger}$: on the gauge-invariant space the two orderings differ by the sign of the matrix elements that create or annihilate a rishon pair, and using $\zeta\zeta^{\dagger}$ at all four corners does \emph{not} give an equivalent Hamiltonian once matter is present. The corner operators are explicitly
\begin{align}
  \hat C_{+x,+y}&=\sum_{c}(\hat\zeta_{c}P)_{+x}\,(\hat\zeta^{\dagger}_{c})_{+y},&
  \tilde C_{-x,+y}&=\sum_{c}(\hat\zeta^{\dagger}_{c}P)_{-x}\,P_{-y}\,P_{+x}\,(\hat\zeta_{c})_{+y},\nonumber\\
  \tilde C_{+x,-y}&=\sum_{c}(P\hat\zeta_{c})_{-y}\,(\hat\zeta^{\dagger}_{c})_{+x},&
  \tilde C_{-y,-x}&=\sum_{c}(P\hat\zeta_{c})_{-x}\,(\hat\zeta^{\dagger}_{c})_{-y},
\label{eq:corners}
\end{align}
which is the form implemented in our code. On the flux-matched, gauge-invariant space, Eq.~\eqref{eq:plaq} coincides with $\Tr(U_1U_2U_3^{\dagger}U_4^{\dagger})$ evaluated with the bosonic link matrices of Eq.~\eqref{eq:Umatrix}; its Hermitian part $\frac12(\Tr\square+{\rm H.c.})$ has eigenvalues in $[-1,1]$ on a single plaquette, the value $1$ being attained by the flux-ordered state.}

Finally, we split the link Casimir symmetrically between its two boundary sites,
\begin{equation}
  \hat\Gamma_{\mathbf j}=\tfrac12\sum_{\ell}\hat T^{2}_{\ell}(\mathbf j),
  \qquad
  \sum_{\mathbf j}\hat\Gamma_{\mathbf j}=\sum_{\rm links}\hat E^{2},
\end{equation}
where the factor of $\tfrac12$ corrects for double counting, as each link is visited once from each of its two endpoints.
Thus, the entire Hamiltonian given in Eq.~\eqref{eq:Hdim_main} takes the form: 
\begin{align}
    H &= m\sum _{\mathbf{j}} \epsj N_j 
       + \frac{1}{2}\sum_{\mathbf{j}} \bigg(-iQ^\dagger_{\boldsymbol{\mu}_x}(\mathbf{j})Q_{-\boldsymbol{\mu}_x}(\mathbf{j}+\boldsymbol{\mu}_x) + \text{H.c.}\bigg) - \frac{1}{2}\sum_{\mathbf{j}}\bigg(\epsj Q^\dagger_{\boldsymbol{\mu}_y}(\mathbf{j})Q_{-\boldsymbol{\mu}_y}(\mathbf{j}+\boldsymbol{\mu}_y) + \text{H.c.} \bigg) \nonumber \\
      &\quad + \frac{g^2}{2} \sum_{\mathbf{j}} \Gamma_{\mathbf{j}} 
       - \frac{1}{2g^2}\sum_{\Box}\CHANGE{\Big(\Tr\square + \text{H.c.}\Big)},
\label{eq:Hdressed}
\end{align}
where $\epsj=(-1)^{j_x+j_y}$ is the staggered phase factor \CHANGE{and $\Tr\square$ is given by Eqs.~\eqref{eq:plaq}--\eqref{eq:corners}. The relative sign between the hopping phases and the magnetic term is physical.}

\section{Gauge singlet condition}

In the hardcore gluon approximation, we truncate the local link Hilbert space
at $j_{\max}=\tfrac{1}{2}$, so every rishon (half-link) carries either the
gauge singlet $\mathbf{1}$ ($j=0$) or the fundamental doublet $\mathbf{2}$
($j=\tfrac{1}{2}$),
\[
  \mathcal{H}_{\text{rishon}} \;=\; \mathbf{1}\oplus\mathbf{2}. 
\]
Physical states must satisfy Gauss's law at every site: the representations of
all objects meeting at a site must combine into the trivial (singlet)
representation of the local $\mathrm{SU}(2)$ gauge group. The number of
gauge-invariant states at a site is therefore the multiplicity of $\mathbf{1}$
in the tensor product of all on-site degrees of freedom. 

Let us first consider the case of pure gauge theory, neglecting the matter fields (considered later). Each of the
links attached to a site contributes one rishon at that site, so a site with coordination number $n_{\text{links}}$ carries $n_{\text{links}}$ rishons. For a bulk site in $d$ spatial dimensions, we have $2d$ rishons, whereas edge and corner sites carry fewer --- this is why the number of singlets depends on both the spatial dimension and the location of
the site. The local gauge Hilbert space of a bulk site is thus
\[
  \left(\mathbf{1}\oplus\mathbf{2}\right)^{\otimes 2d},
\]
and each additional spatial dimension adds two rishons.
Expanding the local Hilbert space,
\[
  \left(\mathbf{1}\oplus\mathbf{2}\right)^{\otimes 2d}
  =
  \underbrace{\left(\mathbf{1}\oplus\mathbf{2}\right)
  \otimes \cdots \otimes
  \left(\mathbf{1}\oplus\mathbf{2}\right)}_{2d\ \text{factors}},
\]
each term selects a subset of the $2d$ rishons to be $\mathbf{2}$ and the rest
$\mathbf{1}$. The term in which every factor is $\mathbf{1}$ gives a single
singlet. More generally, a singlet can arise only from an \emph{even} number of
$\mathbf{2}$ factors, since $\mathbf{2}^{\otimes(\text{odd})}$ carries
half-integer total spin and hence never contains $\mathbf{1}$. The choice of exactly $2k$ of the $2d$ rishons to be $\mathbf{2}$ can be done in 
\[
  \binom{2d}{2k}
\]
ways, and each such choice contributes a number of singlets equal to the
multiplicity of $\mathbf{1}$ in $\mathbf{2}^{\otimes 2k}$. Denoting this
multiplicity by $n_{2k}$, the total number of gauge-invariant singlets is
\[
  N_{\mathrm{singlet}}
  = \sum_{k=0}^{d} \binom{2d}{2k}\, n_{2k}.
\]

The remaining task is to determine $n_{2k}$. We count states of $2k$
spins-$\tfrac{1}{2}$ at fixed total magnetization $M$, the eigenvalue of the
total $S_z$. Because $2k$ is even, every multiplet has integer total spin and
therefore contains exactly one $M=0$ state; the number of $M=0$ states,
$\binom{2k}{k}$ (i.e., $k$ down spins out of $2k$), thus equals the total number
of multiplets. Repeating the count at $M=1$ requires $k-1$ down spins and gives
$\binom{2k}{k-1}$; only multiplets with $j\ge 1$ have a state at this
magnetization (again one each), so this counts the non-singlet multiplets. Their
difference is the number of singlets,
\begin{equation}
  n_{2k}=\binom{2k}{k}-\binom{2k}{k-1}=\frac{1}{k+1}\binom{2k}{k},
  \label{eq:catalan}
\end{equation}
the $k$-th Catalan number. Consequently,
\[
  N_{\mathrm{singlet}}
  = \sum_{k=0}^{d} \binom{2d}{2k}\,\frac{1}{k+1}\binom{2k}{k}.
\]
For example, for $d=2$ the pure gauge theory gives
$1+\binom{4}{2}\cdot 1+\binom{4}{4}\cdot 2 = 9$ physical states per site.
We can now include dynamical fermionic matter. Gauss's law at a site then constrains the \emph{full} on-site product $(\text{rishons})\otimes(\text{matter})$ to
contain the singlet: the $2d$ rishons meeting at the site must be combined with
the on-site matter charge, not merely among themselves. For an $\mathrm{SU}(2)$
fundamental (staggered) fermion, the on-site matter Hilbert space decomposes as
\[
  \mathbf{1}\oplus\mathbf{2}\oplus\mathbf{1},
\]
where the empty and doubly-occupied states are gauge singlets and the
singly-occupied states form a color doublet $\mathbf{2}$.

The two matter singlets attach to every gauge-sector singlet, contributing an
overall factor of two to the pure-gauge count. The matter doublet, on the other
hand, forms a gauge-invariant singlet only when paired with a gauge-sector
doublet, via $\mathbf{2}\otimes\mathbf{2}=\mathbf{1}\oplus\mathbf{3}$; each gauge
doublet then yields exactly one additional singlet. We therefore also need the
number of doublets $\mathbf{2}$ contained in
$\left(\mathbf{1}\oplus\mathbf{2}\right)^{\otimes 2d}$. A doublet requires an
\emph{odd} number $2k+1$ of $\mathbf{2}$ rishons; choosing them in
$\binom{2d}{2k+1}$ ways, each contributing the multiplicity $n_{2k+1}$ of
$\mathbf{2}$ in $\mathbf{2}^{\otimes(2k+1)}$, gives
\[
  \sum_{k=0}^{d-1} \binom{2d}{2k+1}\, n_{2k+1},
  \qquad
  n_{2k+1}=\binom{2k+1}{k}-\binom{2k+1}{k-1},
\]
obtained by the same $M$-counting at $M=\tfrac{1}{2}$ versus $M=\tfrac{3}{2}$.
(Both multiplicities are the same ballot number,
$n_{m}=\binom{m}{\lfloor m/2\rfloor}-\binom{m}{\lfloor m/2\rfloor-1}$, counting
the minimal irrep in $\mathbf{2}^{\otimes m}$: $\mathbf{1}$ for even $m$,
$\mathbf{2}$ for odd $m$.) The total number of on-site gauge-invariant states is
therefore
\begin{equation}
  N_{\mathrm{singlet}}
  = 2\sum_{k=0}^{d} \binom{2d}{2k}\, n_{2k}
  + \sum_{k=0}^{d-1} \binom{2d}{2k+1}\, n_{2k+1},
  \label{eq:algebra_counting}
\end{equation}
leading to 30 gauge singlets per dressed site (in the bulk) for $d=2$. 

\vspace{2mm}
\noindent The singlet counting problem can also be mapped to a geometrical
problem of counting non-intersecting chords between $n$
marked points on a circle. Let $f(n)$ denote the number of
$\SUtwo$ singlets in the tensor product
$(\textbf{1}\oplus\textbf{2})^{\otimes n}$.
Of the $n$ copies, suppose $k$ are ``active'' (contributing
$\textbf{2}$, i.e., $j=\half$) and the remaining $n-k$ are
``inactive'' ($\textbf{1}$, i.e., $j=0$).
There are $\binom{n}{k}$ ways to choose which copies are active. The $k$ active doublets couple to total spin zero only when $k$ is even: a singlet requires the total magnetic quantum number $M = \sum_{i=1}^{k} m_i = 0$ with each $m_i = \pm\half$, so exactly $k/2$ must have $m_i = +\half$ and $k/2$ must have $m_i = -\half$, which forces $k = 2m$. The number of independent singlets in
$\textbf{2}^{\otimes 2m}$ is the Catalan number
$C_m = \binom{2m}{m}/(m+1)$. To see this, couple the $2m$ doublets sequentially: at each step the running angular momentum $j$ changes by $\pm\half$, starting and ending at
$j = 0$, with $j \geq 0$ at every intermediate stage. Each
distinct sequence of intermediate $j$ values defines a
linearly independent recoupling channel. Such sequences are
in bijection with Dyck paths of $2m$ steps, counted by
$C_m$. Explicitly: $C_1 = 1$ (from
$\textbf{2}\otimes\textbf{2} = \textbf{1}\oplus\textbf{3}$, one singlet),
$C_2 = 2$ (from $\textbf{2}^{\otimes 4}$, two singlets via
intermediate sequences
$0 \to \half \to 0 \to \half \to 0$ and
$0 \to \half \to 1 \to \half \to 0$), and so on.
Therefore,
\begin{equation}
  f(n) = \sum_{m=0}^{\lfloor n/2\rfloor}
    \binom{n}{2m}\, C_m
    \;=\; M_n,
  \label{eq:motzkin_def}
\end{equation}
where $M_n$ is the $n$-th Motzkin number~\cite{Motzkin1948}.
The matter sector decomposes as
$\textbf{1}\oplus\textbf{2}\oplus\textbf{1}
= (\textbf{1}\oplus\textbf{2}) \oplus \textbf{1}$.
We count singlets separately for each matter occupation:
\begin{itemize}
\item \textit{$n_M = 0$ or $2$ \;($j_M = 0$):}
  The matter contributes an inert singlet. The $2d$ rishon
  modes must form a singlet on their own, giving
  $f(2d) = M_{2d}$ states. Two such matter sectors
  contribute $2\,M_{2d}$.

\item \textit{$n_M = 1$ \;($j_M = \half$):}
  The matter contributes a doublet. Together with the $2d$
  rishons, the full system is
  $\textbf{2}\otimes(\textbf{1}\oplus\textbf{2})^{\otimes 2d}$.
  Since the matter doublet is always active
  ($j_M = \half \neq 0$), its contribution is the number of
  singlets in $(\textbf{1}\oplus\textbf{2})^{\otimes(2d+1)}$ with
  the constraint that the matter copy is active, which equals
  $f(2d{+}1) - f(2d) = M_{2d+1} - M_{2d}$.
\end{itemize}
Adding all three sectors, the total gauge-invariant
dressed-site dimension in $d$ spatial dimensions is
\begin{equation}
  D(d) = 2\,M_{2d} + (M_{2d+1} - M_{2d})
       = M_{2d} + M_{2d+1}\,.
  \label{eq:geometry_counting}
\end{equation}
It is straightforward to check that Eqs.~\eqref{eq:algebra_counting} and \eqref{eq:geometry_counting} are equivalent counting expressions. 
Motzkin numbers $M_n$ with $n = 0, 1, 2 \cdots$ form the sequence: $1, 1, 2, 4, 9, 21, 51, 127, 323, 835, \cdots$. For example, $D(2) = M_4 + M_5 = 30$ and $D(1) = M_2 + M_3 = 6$ as found in Ref.~\cite{Jha:2026ror}, while
$D(3) = M_6 + M_7 = 178$ and quantum simulation of $\SUtwo$ gauge theory in the hardcore gluon truncation will require at most $8L^3$ qubits for a cubic lattice of size $L^3$ with open boundary conditions \CHANGE{(eight qubits per bulk site, fewer on the boundary)}.

\section{Qubit encoding of the dressed sites}
\label{app:gauge-singlet}
Together with the fermionic rishon field, the raw on-site Hilbert space has dimension: 
\begin{equation}
    \dim \mathcal{H}^{\mathrm{raw}}_{\mathrm{site}}
    =
    4 \times 3^{n_{\mathrm{links}}}
    =
    \begin{cases}
        4 \times 3^4 = 324,
        & \text{bulk } (n_{\mathrm{links}}=4), \\[4pt]
        4 \times 3^3 = 108,
        & \text{edge } (n_{\mathrm{links}}=3), \\[4pt]
        4 \times 3^2 = 36,
        & \text{corner } (n_{\mathrm{links}}=2).
    \end{cases}
    \label{eq:raw_space_counting}
\end{equation}
This entire Hilbert space is strongly redundant, because it contains states that violate
the non-Abelian Gauss law. In the dressed-site representation, Gauss's law
becomes a strictly \emph{on-site} constraint: physical states must be color
singlets under the total generator of matter and rishons,
\begin{equation}
    \hat{G}^{a}_{\mathbf{j}}
    =
    \hat{S}^{a}_{\mathrm{matter}}(\mathbf{j})
    +
    \sum_{\ell}
    \hat{T}^{a}_{\ell}(\mathbf{j}),
    \qquad
    a = 1,2,3 .
\end{equation}
Equivalently, physical states satisfy
\begin{equation}
    \hat{G}^{2}_{\mathbf{j}}
    \ket{\mathrm{phys}}
    =
    \left[
        \left(\hat{G}^{1}_{\mathbf{j}}\right)^2
        +
        \left(\hat{G}^{2}_{\mathbf{j}}\right)^2
        +
        \left(\hat{G}^{3}_{\mathbf{j}}\right)^2
    \right]
    \ket{\mathrm{phys}}
    =
    0 .
\end{equation}
Projecting onto the gauge-invariant sector reduces the physical dimension to: 
\begin{equation}
    \dim \mathcal{H}^{\mathrm{phys}}_{\mathrm{site}}
    =
    \begin{cases}
        30, & \text{bulk}, \\[4pt]
        13, & \text{edge}, \\[4pt]
        6,  & \text{corner}, 
    \end{cases}
    \label{eq:dressed_physical_count}
\end{equation}
which is only $9\%, 12\%, 16.7\%$ of the full Hilbert space of Eq.~\eqref{eq:raw_space_counting}
for bulk, edge, and corner, respectively. Because the constraint in Eq.~\eqref{eq:gauss} is local, it can be imposed once
when the on-site basis is constructed, after which the Hamiltonian acts entirely
within the physical space and no further gauge projection is required. Only the link condition, which couples neighboring sites, remains to be enforced. The total qubit count is:
\begin{equation}
N_q = 3N_{\rm corner} + 4N_{\rm edge} + 5N_{\rm bulk},
\end{equation}
and using $N_{\rm corner}=4$,
$N_{\rm edge}=2(L_x-2)+2(L_y-2)$,
$N_{\rm bulk}=(L_x-2)(L_y-2)$, we find:
\begin{equation}
N_q = 5L_xL_y - 2L_x - 2L_y
\end{equation}
for a lattice of size $L_x \times L_y$. The maximum lattice size we consider in this work is \CHANGE{$8 \times 8$ corresponding to 288 qubits}. In addition, we focus exclusively on the zero baryon $(b=0)$ sector, which leads to a further reduction of the dimension of the Hilbert space; \CHANGE{this reduction is only polynomial in the volume, as opposed to the exponential one obtained by passing from the raw space, Eq.~\eqref{eq:raw_space_counting}, to the physical dressed space, Eq.~\eqref{eq:dressed_physical_count}.}

\section{Stabilizers, SRE, and non-local SRE}

\subsection{Stabilizer formalism}

The stabilizer formalism has become one of the fundamental frameworks in quantum information science due to its versatility for constructing quantum error correcting codes.
For an $n$-qubit system, the Clifford group is defined as the set of unitary operations that preserve the Pauli group under conjugation and can be generated by the Hadamard, S, and CNOT gates. Quantum states that are reachable from a simple product state using only Clifford operations are known as \emph{stabilizer states}. Although these states exhibit highly nontrivial quantum features, including the ability to support maximal entanglement, they remain efficiently simulable on classical computers according to the Gottesman--Knill theorem~\cite{Gottesman:1998hu}. This observation establishes that stabilizer operations alone cannot provide a computational advantage, making non-stabilizer resources an unavoidable ingredient for universal quantum computation~\cite{BravyiKitaev:2005, Veitch:2014} and for evaluating the scope beyond classical computation.  Despite its central importance, characterizing and quantifying magic remains a challenging task. Most of the promising magic measures become extremely challenging to evaluate for large many-body systems~\cite{Leone:2021rzd}. This restricted both analytical and numerical investigations primarily to systems with relatively few degrees of freedom~\cite{Tarabunga:2024}. In parallel, growing interest has focused on understanding quantum many-body phases from the perspective of computational complexity. 

Rather than characterizing phases solely through conventional order parameters, this approach seeks to understand the computational resources required to prepare, manipulate, or simulate quantum states. Such ideas have established connections between magic and a broad range of problems in quantum information, including the emergence of sign problems in quantum Monte Carlo methods, and the computational capabilities of many-body quantum states~\cite{Ellison:2021,Heinrich:2019,Tarabunga:2024}. These developments highlight that magic is not only a resource for fault-tolerant quantum computation, but also a powerful probe of the intrinsic computational complexity underlying quantum many-body systems.

The basic algebraic object is the $n$-qubit Pauli group, denoted by $\mathcal{P}_n$.
It consists of all tensor products of the single-qubit Pauli operators: 
\[
\mathbb{I}=\begin{pmatrix}1&0\\0&1\end{pmatrix}, \qquad
X=\begin{pmatrix}0&1\\1&0\end{pmatrix}, \qquad
Y=\begin{pmatrix}0&-i\\i&0\end{pmatrix}, \qquad
Z=\begin{pmatrix}1&0\\0&-1\end{pmatrix},
\]
together with the global phases \(\pm1\) and \(\pm i\). Explicitly,
\[
\mathcal{P}_n=\left\{
\alpha\, P_1\otimes P_2\otimes\cdots\otimes P_n
\,\middle|\,
P_i\in\{\mathbb{I},X,Y,Z\},
\ \alpha\in\{\pm1,\pm i\}
\right\}.
\]
The Pauli group provides a complete operator basis for the space of linear operators on the \(2^n\)-dimensional Hilbert space. 

The Clifford group $C_n$ is defined as a set of operators that map the set of \textit{n}-fold Pauli group products into itself. 
In other words, the Clifford group preserves the algebraic structure of the Pauli group under unitary transformations. The entire Clifford group can be generated using only three elementary quantum gates: the Hadamard gate \(H\), the phase gate \(S\), and the controlled-NOT (CNOT) gate. For example, the Hadamard gate satisfies
\[
HXH=Z,\qquad
HZH=X,\qquad
HYH=-Y,
\]
illustrating that conjugation by a Clifford operator simply permutes the Pauli operators among themselves. The action of the Clifford group naturally defines the class of stabilizer states. Beginning with the trivial computational basis state $|0\rangle^{\otimes n}$, every pure stabilizer state can be written as
$|\psi\rangle = U|0\rangle^{\otimes n}$ for some $U\in C_n$. 
Thus, the set of pure stabilizer states consists precisely of all states that are reachable from the computational basis using Clifford operations. These states possess many genuinely quantum properties, including multipartite entanglement, yet they remain efficiently simulable on a classical computer. This remarkable property is a direct consequence of the Gottesman--Knill theorem, which states that any quantum circuit composed entirely of Clifford gates acting on stabilizer states can be simulated in polynomial time on a classical computer. Consequently, stabilizer states are regarded as the free states within the resource theory of magic, while quantum computational advantage requires states that lie outside this class. 

Realistic quantum systems are generally described by mixed states rather than pure states. The set of all mixed stabilizer states is obtained by taking probabilistic mixtures of pure stabilizer states. Mathematically, this set is defined as the convex hull of all pure stabilizer states,
\[
\mathrm{STAB}
=
\mathrm{conv}
\left\{
|\psi\rangle\langle\psi|
:\,
|\psi\rangle
=
U|0\rangle^{\otimes n},
\;
U\in C_n
\right\}.
\]
Equivalently, every state inside the stabilizer polytope can be expressed as
\[
\rho
=
\sum_i
p_i
|\psi_i\rangle\langle\psi_i|,
\]
where $p_i\ge0$, $\sum_i p_i=1,$ and each \(|\psi_i\rangle\) is a pure stabilizer state. The pure stabilizer states themselves form the extreme points (vertices) of this convex set. Let $S$ denote the set of all $n$-qubit quantum states,
$S
=
\left\{
\rho
:
\rho\ge0,\;
\mathrm{Tr}(\rho)=1
\right\}.$
Since every stabilizer state is a valid quantum state,
the full state space \(S\) forms a convex body whose dimension is $4^n-1.$

Geometrically, \(\mathrm{STAB}\) is therefore a convex polytope embedded inside the much larger convex set \(S\). Unlike a sphere or an ellipsoid, a polytope is completely determined by a finite set of vertices. In the present case, these vertices are precisely the pure stabilizer states. One remarkable feature of this polytope is the enormous number of its vertices. For an \(n\)-qubit system, the number of pure stabilizer states scales super-exponentially with system size.
Consequently, the geometry of the stabilizer polytope becomes increasingly intricate as the number of qubits increases, making many exact calculations of magic measures computationally difficult and intractable beyond a few qubits. This also leads to the superexponential cost of determining the boundary of quantum advantage~\cite{Leone:2026kxv}.  Another important property is the high degree of symmetry exhibited by the stabilizer polytope. Since the Clifford group maps every stabilizer state into another stabilizer state,
\[
U\rho U^\dagger\in\mathrm{STAB},
\qquad
\forall\,
U\in C_n,
\]
the entire polytope is invariant under Clifford transformations. In group-theoretic language, the Clifford group acts multiply transitively on the vertices of the polytope, implying that all pure stabilizer states are equivalent under Clifford operations and that none occupies a distinguished position within the stabilizer set~\cite{Heinrich:2019}. In contrast, any quantum state satisfying
\begin{equation}
 \rho\notin\mathrm{STAB},    
\end{equation}
cannot be represented as a convex combination of stabilizer states and is said to possess quantum magic. Such non-stabilizer states provide the additional resource required for universal fault-tolerant quantum computation and are the primary objects quantified by modern measures of quantum magic.

\subsection{Stabilizer R\'enyi entropy}

Until recently, no easily measurable observable $M (\rho)$ behaving as a true magic monotone was available. Such an observable must satisfy the following resource-theoretic requirements: 
\begin{align}
&\text{(a) } M(\rho) = 0 \text{ for every stabilizer state } \rho; \\
&\text{(b) } M(C\rho C^\dagger) = M(\rho) \text{ for every Clifford unitary } C; \\
&\text{(c) } M(\rho\otimes\rho') = M(\rho) + M(\rho') \quad \text{(additivity)}.
\end{align}
Following the work of Ref.~\cite{Leone:2021rzd}, we now have 
stabilizer R\'enyi entropy (SRE) of order $\alpha$, $M_\alpha(\ket{\psi})$, as a good magic monotone for $\alpha \ge 2$. 
Any density matrix $\rho$ on $n$ qubits can be expanded in the Pauli basis,
\begin{equation}
\rho = \frac{1}{d}\sum_{P\in\mathcal{P}_n} \mathrm{Tr}(P\rho)\, P, \qquad d = 2^n,
\end{equation}
since the $d^2$ Pauli strings form a complete orthogonal operator basis for the space of linear operators on the $d$-dimensional Hilbert space. Writing
$\langle P\rangle \equiv \mathrm{Tr}(P\rho) = \langle\psi|P|\psi\rangle$ for a pure state $\rho=|\psi\rangle\langle\psi|$,
define the normalized Pauli spectrum
\begin{equation}
\Xi_P(\psi) \;=\; \frac{\langle P\rangle^2}{d}.
\label{eq:xi-def}
\end{equation}
A stabilizer state has this Pauli weight concentrated on a small set of Pauli operators; a highly non-stabilizer state spreads it over many. The SRE quantifies precisely this spread. As $P$ is Hermitian and $\langle\psi|P|\psi\rangle \in [-1, 1]$, every $\Xi_P \geq 0$; summing to unity follows directly from purity ($\mathrm{Tr}(\rho^2)=1$), and we get: 
\begin{equation}
\mathrm{Tr}(\rho^2) = \frac{1}{d^2}\mathrm{Tr}\!\left[\left(\sum_P \langle P\rangle P\right)^{\!2}\,\right]
= \frac{1}{d}\sum_P \langle P\rangle^2 = \sum_P \Xi_P = 1,
\end{equation}
using $\mathrm{Tr}(PP') = d\,\delta_{PP'}$. Thus, for any pure state, $\{\Xi_P\}_{P\in\mathcal{P}_n}$ is a probability distribution over the $4^n$ Pauli strings, and we can quantify its spread with a R\'enyi entropy.
Throughout this work, we use the R\'enyi-2 stabilizer entropy $M_{2}$ unlike other members of the more general SRE family
\footnote{Applying $H_{\alpha}(p) = \frac{1}{1-\alpha} \log  p_{i}^{\alpha} $ to $\Xi_{P}$ and subtracting $\log d$ for normalization defines $M_{\alpha}(\psi) = H_{\alpha}(\Xi)-\log d$ for any R\'enyi index $\alpha$; each member vanishes on stabilizer states by the same argument given below for $\alpha=2$, but $M_{2}$ is the only one with a known direct measurement protocol.}.  
Using $\Xi_{P}^2 = \langle P\rangle^4/d^{2}$, we get~\cite{Leone:2021rzd}:
\begin{equation}
M_2(\ket{\psi}) = -\log_2\!\left(\frac{1}{d}\sum_P \langle \psi \vert P \vert \psi\rangle^4\right).
\label{eq:m2-final}
\end{equation}

\paragraph{Vanishing on stabilizer states.} For a pure stabilizer state $|\psi\rangle$, there are exactly $d$ Pauli operators forming its stabilizer group satisfying $P|\psi\rangle = \pm|\psi\rangle$ and hence $|\langle P\rangle| = 1$, giving $\Xi_P = 1/d$ for each of these $d$ operators; for every other Pauli string,
$\Xi_P = 0$ identically. Substituting into Eq.~\eqref{eq:m2-final}, we obtain
\begin{equation}
\frac{1}{d}\sum_P \langle P\rangle^4 = 1, 
\end{equation}
confirming that $M_2$ vanishes exactly on the free states of the resource theory, and is strictly positive whenever the Pauli spectrum spreads beyond this minimal $d$-element support.

\paragraph{Measurability.} The quantity $\langle P\rangle^4$ is nonlinear in $\rho$, so it cannot be written as $\mathrm{Tr}(\rho O)$ for a fixed operator $O$ acting on a single copy of the state. This obstruction is resolved by the replica trick: since
\begin{equation}
\langle P\rangle^4 = \big(\langle\psi|P|\psi\rangle\big)^4 = \langle\psi|^{\otimes 4}\, P^{\otimes 4} \,|\psi\rangle^{\otimes 4},
\end{equation}
the nonlinear single-copy quantity $\langle P\rangle^4$ is exactly equal to a linear expectation value of the fixed observable $P^{\otimes 4}$ on four copies of the state, $\rho^{\otimes 4}$.

\subsection{\label{app:subsec_lowerbound_and_nlmagic}Lower bound and computation of non-local magic}

For many practical purposes, the total magic is both uninformative and expensive
to compute, a problem that becomes especially severe in two dimensions. A more
tractable and physically sharper object is the part of the magic (for a given
bipartition) that \emph{cannot} be removed by local unitaries acting on the two
subsystems. This quantity, the \emph{non-local} magic, has been shown to be
bounded below by the anti-flatness of the entanglement spectrum (up to a
prefactor). For a bipartition with reduced density matrix $\rho\equiv\rho_A$ whose
eigenvalues (the entanglement spectrum) are $\{\lambda_i\}$, we use the
lowest-order \emph{anti-flatness}~\cite{Tirrito:2023fnw}
\begin{equation}
F(\rho)=\Tr\rho^{3}-\big(\Tr\rho^{2}\big)^{2}
       =\sum_i\lambda_i^{3}-\Big(\sum_i\lambda_i^{2}\Big)^{2},
\label{eq:antiflat}
\end{equation}
which vanishes for a flat (stabilizer-like) spectrum. It has been
argued~\cite{Cao:2024nrx, Liu:2026twg} that $4F$ is a lower bound on the non-local
stabilizer R\'enyi entropy (SRE) $M_2^{\rm NL}$ across the cut, while
$M_2^{\rm NL}$ is itself bounded above by the full SRE $M_2$. As shown in the next section (App.~\ref{app:proof_of_tighter_bound}), we find a stronger inequality, giving the sandwich
\begin{equation}
-\log_{2}(1-4F) \;\le\; M_2^{\rm NL} \;\le\; M_2 .
\label{eq:sandwich}
\end{equation}
\CHANGE{Combined with the results of Ref.~\cite{Sierant:2026lvg}, this sharpens into the conjecture-free bracket of Eq.~\eqref{eq:bracket}.}
The lower bound of the non-local magic does not require Pauli sampling, and can be evaluated
for very large system sizes\CHANGE{, although it never exceeds $\log_2(64/37)\approx0.79$ bits (see the discussion after Corollary~\ref{cor:bracket})}; on the contrary, $M_2^{\rm NL}$ is defined through a
difficult optimization, and $M_2$ requires sampling an exponentially large Pauli
space. However, recently, Ref.~\cite{Liu:2026nah} proposed a computationally cheap route to
$M_2^{\rm NL}$, which we adopt. The non-local SRE of a bipartition is the magic
that cannot be removed by any local unitary on either side,
\begin{equation}
M_2^{\rm NL}(\ket\psi)=\min_{U_A\otimes U_B} M_2\big((U_A\otimes U_B)\ket\psi\big).
\label{eq:nldef}
\end{equation}
Building on Refs.~\cite{Tarabunga:2023xmv, Cao:2024nrx}, Ref.~\cite{Liu:2026nah}
evaluates Eq.~\eqref{eq:nldef} at a cut through a Schmidt-reference-state
construction. Order the entanglement spectrum as
$\lambda_0\ge\lambda_1\ge\dots\ge\lambda_{\chi-1}>0$ (with $\sum_i\lambda_i=1$ and
$\chi$ the Schmidt rank), and define the reference state on
$n_{\rm ref}=2\lceil\log_2\chi\rceil$ qubits ($m=\lceil\log_2\chi\rceil$ per side), as the state
\begin{equation}
|\tilde{\psi}\rangle=\sum_{i=0}^{\chi-1}\sqrt{\lambda_i}\,
\ket{i}_{\widetilde  A}\ket{i}_{\widetilde  B},
\end{equation}
i.e., a computational-basis encoding of the ordered spectrum along the diagonal
$\ket{i}\ket{i}$. Among all assignments of the $\lambda_i$ to the labels
$\ket{i}$, the descending ordering \CHANGE{is conjectured to yield} the smallest SRE of $|\tilde{\psi}\rangle$,
and hence the \emph{tightest} upper bound \CHANGE{of this form} on the non-local SRE~\cite{Liu:2026nah}.
The SRE of the reference state has the closed form
\begin{equation}
M_2(|\tilde{\psi}\rangle)
=-\log_2\!\!\sum_{i_1i_2i_3i_4}\!\!
\sqrt{\lambda_{i_1}\lambda_{i_2}\lambda_{i_3}\lambda_{i_4}}\,
\sqrt{\lambda_{i_3\oplus i_2\oplus i_1}\lambda_{i_4\oplus i_2\oplus i_1}
      \lambda_{i_1\oplus i_3\oplus i_4}\lambda_{i_2\oplus i_3\oplus i_4}},
\label{eq:nlref}
\end{equation}
with $\oplus$ denoting the bitwise XOR. Equation~\eqref{eq:nlref} is \CHANGE{a rigorous upper
bound, $M_2^{\rm NL}\le M_2(|\tilde{\psi}\rangle)$, without any conjecture (see Corollary~\ref{cor:bracket})}. Ref.~\cite{Liu:2026nah} further
conjectures that the descending reference state Schmidt coefficients
results in the smallest value of Eq.~\eqref{eq:nlref} and leads through the conjecture of Ref.~\cite{Liu:2026nah} to $M_2^{\rm NL}=M_2(|\tilde{\psi}\rangle)$. \CHANGE{The conjecture has been proven for Schmidt rank $\chi\le6$ and for dyadic-staircase spectra, and in general $M_2(|\tilde\psi\rangle)-4\le M_2^{\rm NL}\le M_2(|\tilde\psi\rangle)$~\cite{Sierant:2026lvg}.} We adopt \CHANGE{$M_2(|\tilde\psi\rangle)$} as our estimate of the non-local magic.

In practice, we construct $|\tilde{\psi}\rangle$ at the middle cut and evaluate
Eq.~\eqref{eq:nlref} with the same fast Walsh--Hadamard transform (FWHT) used for
the exact $M_2$\CHANGE{; see Ref.~\cite{Sierant:2026jru} for FWHT-based algorithms for the SRE of state vectors}. Because $|\tilde{\psi}\rangle$ lives on only
$n_{\rm ref}=2\lceil\log_2\chi\rceil$ qubits, this is inexpensive even in two
dimensions if $\chi$ is not very large.
\CHANGE{Since our estimate of $M_2^{\rm NL}$ is $M_2(|\tilde\psi\rangle)$, the fact that it lies above $-\log_2(1-4F)$ for all lattice volumes and all $g^2$ is guaranteed by Theorem~\ref{thm:main} together with Corollary~\ref{cor:bracket}. } 
\section{\label{app:proof_of_tighter_bound}Proof of improved inequality, Eq.~\eqref{eq:strong_inequality}}

Consider a normalized pure state $\ket{\psi}$ on two registers of $n_A$ and $n_B$
qubits, of dimensions $d_A=2^{n_A}$ and $d_B=2^{n_B}$.
Let $\rho_A=\Tr_B|\psi\rangle\langle\psi|$, and define the anti-flatness and stabilizer purity as: 
\[
 F=\Tr\rho_A^3-(\Tr\rho_A^2)^2,\qquad
 W=\frac1{d_Ad_B}\sum_{P,Q}\langle\psi|P\otimes Q|\psi\rangle^4,
\]
where $P$ and $Q$ run over the $d_A^2$ and $d_B^2$ tensor products of Pauli matrices
$\mathbb{I},X,Y,Z$ on their respective registers, with no
additional phases, so that $P\otimes Q$ runs over all Hermitian Pauli
strings of the joint system and every expectation value is real.
There is no restriction on the rank or spectrum of the mixed state $\rho_A$.
Note that $F\ge0$, since by Cauchy--Schwarz,
$\bigl(\Tr\rho_A^2\bigr)^2=\bigl(\Tr\rho_A^{1/2}\rho_A^{3/2}\bigr)^2
\le\bigl(\Tr\rho_A\bigr)\bigl(\Tr\rho_A^3\bigr)=\Tr\rho_A^3$. 

\begin{theorem}\label{thm:main}
With the above notation, we have $4F\le1-W$. The coefficient $4$ is optimal.
\end{theorem}

Recall that the linear and R\'enyi-2 stabilizer entropies are
$M_{\rm lin}=1-W$ and $M_2=-\log_2W$ (see~\cite{Leone:2021rzd}).
We can now derive Theorem~\ref{thm:sandwich} from the main text, which is
restated as the following corollary.

\begin{corollary}\label{cor:log}
For every pure $\psi$ and every bipartition, we have
$W\ge(d_Ad_B)^{-1}>0$ and
\[
 M_{\rm lin}^{\rm NL}(\psi)\ \ge\ 4F,\qquad
M_2^{\rm NL}(\psi)\ \ge\ -\log_2(1-4F).
\]
Since $F$ depends only on the Schmidt spectrum, minimizing over local
unitaries gives
$M_{\rm lin}^{\rm NL}\ge4F$ and $M_2^{\rm NL}\ge-\log_2(1-4F)$,
for arbitrary Schmidt rank and without assuming that the
computational-basis representative is optimal.
\end{corollary}
\begin{proof}
The term $P=Q=I$ alone contributes $(d_Ad_B)^{-1}$ to $W$, and all terms
are nonnegative. Theorem~\ref{thm:main} is the first inequality, and
gives $0<W\le1-4F$, so $M_2=-\log_2W\ge-\log_2(1-4F)$. Both $M_{\rm lin}$
and $M_2$ are then bounded below by a function of $F$ alone, which is
invariant under $U_A\otimes U_B$.
\end{proof}

The rest of this appendix is devoted to the proof of Theorem~\ref{thm:main}.
The proof will be based on the five lemmas below. First, let us introduce some notation.
For a Hermitian matrix $X$ and a density matrix $\sigma$, define
\begin{align}
 \PP(X)&=\Tr (X^2)(\Tr|X|)^2-2\Tr|X|\Tr|X|^3+2(\Tr X^2)^2,\label{eq:P}\\
 \QQ_\sigma(X)&=\Tr (X^2)+2\Tr(\sigma^2)\Tr(\sigma X^2)
             -4\Tr(\sigma^2X^2)+2\Tr(\sigma X\sigma X).\label{eq:Q}
\end{align}
In terms of the eigenvalues $x_i$ of $X$ and $u_i=|x_i|$, $t=\sum_iu_i^2$, we have
\begin{equation}\label{eq:Pspec}
 \PP(X)=\sum_{i,j}u_iu_j\bigl(t-(u_i-u_j)^2\bigr).
\end{equation}
Note that here $X$ denotes an arbitrary Hermitian matrix and not the Pauli-$X$ matrix.

\begin{lemma}\label{lem:pauli}
Let $X$ be a Hermitian matrix on a $d$-dimensional qubit register, and let $Q$ run
over the $d^2$ Pauli strings of that register. Then
\[\displaystyle d^{-1}\sum_Q\bigl(\Tr(XQ)\bigr)^4\le\PP(X).\]
\end{lemma}
\begin{proof}
Because the Pauli strings are orthonormal, $\Tr(PP')=d\,\delta_{PP'}$, we have the expansion
\[
X=d^{-1}\sum_Pr_PP, \qquad r_P=\Tr(XP).
\]
Writing $QPQ=\chi(P,Q)P$ with $\chi(P,Q)=\pm1$, we also have orthogonality of characters:
\[
\sum_Q\chi(P,Q)\chi(P',Q)=d^2\delta_{PP'}.
\]
From here, we derive
\[
 \Tr(XQXQ)=d^{-1}\sum_P\chi(P,Q)r_P^2,
 \qquad \sum_Q\bigl(\Tr(XQXQ)\bigr)^2=\sum_Qr_Q^4.
\]

Fix an orthonormal eigenbasis $\{\ket{i}\}$ of $X$, let $x_i$ be the corresponding eigenvalues,
$u_i=|x_i|$, and $t=\sum_i u_i^2=\Tr X^2$. Introduce the matrix $m(Q)$ with entries
and $m_{ij}(Q)=|\langle i|Q|j\rangle|^2$. Since $Q$ is unitary,
$m(Q)$ is doubly stochastic, and $\sum_Qm_{ij}(Q)=d$ for all $i,j$
because $\sum_QQ\,M\,Q=d\,\Tr(M)\,I$. As $Q$ is also Hermitian, we have
\[
 \Tr(XQXQ)=\sum_{i,j}x_ix_j\,m_{ij}(Q).
\]
Applying Cauchy--Schwarz to the splitting
$x_ix_jm_{ij}=(w_{ij}m_{ij})^{1/2}\cdot(x_i^2x_j^2m_{ij}/w_{ij})^{1/2}$
with the weights $w_{ij}=(u_i^2+u_j^2)/2$, and using that
$\sum_{i,j}w_{ij}m_{ij}(Q)=t$ by double stochasticity, we obtain
\[
 \bigl(\Tr(XQXQ)\bigr)^2\ \le\ t\sum_{i,j}\frac{2u_i^2u_j^2}{u_i^2+u_j^2}\,m_{ij}(Q).
\]
Summing over $Q$, dividing by $d$ and using $\sum_Qm_{ij}(Q)=d$, we get
\[
 d^{-1}\sum_Qr_Q^4\le t\sum_{i,j}\frac{2u_i^2u_j^2}{u_i^2+u_j^2}
 \le\sum_{i,j}u_iu_j\bigl(t-(u_i-u_j)^2\bigr)=\PP(X),
\]
by \eqref{eq:Pspec}. Terms with $u_i=u_j=0$ are read as zero on both sides.
The second inequality holds termwise: for $u_i^2+u_j^2>0$ the difference is
\[
 u_iu_j\Bigl(t-(u_i-u_j)^2-\frac{2tu_iu_j}{u_i^2+u_j^2}\Bigr)
 =u_iu_j(u_i-u_j)^2\Bigl(\frac{t}{u_i^2+u_j^2}-1\Bigr)\ge0,
\]
since $t\ge u_i^2+u_j^2$; for $i=j$ it is zero. This proves the lemma.
\end{proof}

\begin{lemma}\label{lem:elementary1}
For a positive-semidefinite matrix $G\succeq0$ and arbitrary vectors $u,v$, we have
\begin{equation}\label{eq:sharp}
 \langle u,Gu\rangle\langle v,G^3v\rangle
 +\langle v,Gv\rangle\langle u,G^3u\rangle
 \le\Tr (G^2)\langle u,Gu\rangle\langle v,Gv\rangle
       +|\langle u,G^2v\rangle|^2.
\end{equation}
\end{lemma}
\begin{proof}
First, note that zero values of $\langle u,Gu\rangle$ or
$\langle v,Gv\rangle$ make the claim immediate. Otherwise, normalize
$p=G^{1/2}u/\sqrt{\langle u,Gu\rangle}$
and $q=G^{1/2}v/\sqrt{\langle v,Gv\rangle}$, which are unit vectors.
The identity
\[
 \|(I-pp^*)G(I-qq^*)\|_{\rm HS}^2
 =\Tr G^2-\langle p,G^2p\rangle-\langle q,G^2q\rangle
        +|\langle p,Gq\rangle|^2\ge0
\]
proves the claim after multiplication by
$\langle u,Gu\rangle\langle v,Gv\rangle$, because
$\langle p,G^2p\rangle=\langle u,G^3u\rangle/\langle u,Gu\rangle$ and
$\langle p,Gq\rangle=\langle u,G^2v\rangle/
\sqrt{\langle u,Gu\rangle\langle v,Gv\rangle}$.
\end{proof}

\begin{lemma}\label{lem:elementary2}
If\/ $0\le a,b,c$ and\/
$h(a,b;c)=\max_{0\le s\le1}(2\sqrt{ab}\,s-cs^2)$, then
\begin{equation}\label{eq:three}
 \sum_{\rm cyc}\bigl(h(a,b;c)-(a+b-N)_+\bigr)\le\tfrac32N,
 \qquad x_+=\max(x,0),
\end{equation}
for every $N\ge0$, where the cyclic sum runs over the three assignments
$(a,b;c)$, $(b,c;a)$, $(c,a;b)$. Both sides of \eqref{eq:three} are
symmetric in $a,b,c$, since $h$ is symmetric in its first two arguments.
\end{lemma}
\begin{proof}
Note that the maximum defining $h$ is attained at
$s=\min(1,\sqrt{ab}/c)$, so that $h(a,b;c)=ab/c$ when $ab\le c^2$ and
$h(a,b;c)=2\sqrt{ab}-c$ when $ab\ge c^2$ (and $h=0$ when $a b=c=0$).
By symmetry, we may sort $a\le b\le c$, and we first prove
$H:=\sum_{\rm cyc}h(a,b;c)\le a/2+b+3c/2$.
The cases $b=0$ or $c=0$ force $a=b=0$ and $a=b=c=0$, respectively, and are immediate.
Otherwise, set
$u=\sqrt{b/c}$, $v=\sqrt{a/c}$, so $0\le v\le u\le1$. 

Then
$ab=c^2u^2v^2\le c^2$ and $bc=c^2u^2\ge c^2v^4=a^2$, so the first two
terms are $h(a,b;c)=cu^2v^2$ and $h(b,c;a)=c(2u-v^2)$, while
$h(c,a;b)$ splits according to the sign of $v-u^2$:
\[
 H/c=2u-v^2+u^2v^2+
 \begin{cases}v^2/u^2,&v\le u^2,\\2v-u^2,&v\ge u^2.\end{cases}
\]
If $v\le u^2$, then $H/c-(\tfrac32+u^2+\tfrac12v^2)$ is affine in $v^2$
with slope $u^2+u^{-2}-\tfrac32>0$, hence is largest at $v^2=u^4$, where
\[
 H/c-(\tfrac32+u^2+\tfrac12v^2)
 \le u^6-\tfrac32u^4+2u-\tfrac32
 \le-\tfrac12u^4+2u-\tfrac32
 =-\tfrac12(1-u)^2(u^2+2u+3)\le0,
\]
the middle step being $u^6\le u^4$.
If $v\ge u^2$, the opposite difference is equal to
\[
 (2-v^2)\left(u-\frac1{2-v^2}\right)^2
 +\frac{(1-v)^2(2-v^2/2)+v^3(1-v)}{2-v^2}\ge0,
\]
with each term nonnegative for $0\le v\le1$.
Finally, since $(x)_+\ge x$ and $(x)_+\ge\tfrac12x$,
$\sum_{\rm cyc}(a+b-N)_+\ge\tfrac12(a+c-N)+(b+c-N)
=a/2+b+3c/2-3N/2$, proving \eqref{eq:three}.
\end{proof}

\begin{lemma}\label{lem:operator}
For two positive-semidefinite matrices $A,B\succeq0$ of the same size, put \[N_j=\Tr(A^j+B^j), \qquad
\Gamma=N_1^2N_2+N_1N_3-N_2^2-N_4, \qquad
H=A+B, \qquad M=A^3+B^3, \qquad D=A^2-B^2.\] Then
\begin{equation}\label{eq:Z}
 Z(A,B):=\Gamma H+N_2M+2N_3H^2-2N_1(HM+MH)+2N_1D^2\succeq0.
\end{equation}
\end{lemma}
\begin{proof}
We first prove a scalar matrix inequality. Given $g_i\ge0$, $i=1,\dots,m$, set
\[s_j=\sum_{i=1}^m g_i^j, \qquad c=s_1^2s_2+s_1s_3-s_2^2-s_4,\] and define the $m\times m$ matrix $\K(g)$ with entries
\begin{equation}\label{eq:kernel}
 \begin{split}
 \K(g)_{ii}&=cg_i+s_2g_i^3+2s_3g_i^2-2s_1g_i^4,\\
 \K(g)_{ij}&=2g_ig_j\bigl(s_3-s_1(g_i^2+g_ig_j+g_j^2)\bigr),
 \qquad i\ne j.
 \end{split}
\end{equation}

We will prove by induction on $m$ that $\K(g)\succeq0$. For one entry $(m=1)$, this is $\K(g)=(g_1^5)$.
For two entries $(m=2)$, writing $x=g_1$, $y=g_2$, we have
\[
 \K(x,y)=\diag(x,y)
 \begin{pmatrix}
 x^3+x^2y+xy^2+5y^3&-4xy(x+y)\\
 -4xy(x+y)&y^3+y^2x+yx^2+5x^3
 \end{pmatrix}\diag(x,y),
\]
whose middle matrix has a nonnegative diagonal and determinant
$(x-y)^2(5x^4+16x^3y+18x^2y^2+16xy^3+5y^4)\ge0$.

For the induction step, we prove for $m\ge2$ and $g_i\ge t\ge0$ that
\begin{equation}\label{eq:increment}
 \K(g_1,\ldots,g_m,t)-\bigl(\K(g_1,\ldots,g_m)\oplus0\bigr)\succeq0.
\end{equation}
Write $g_i=h_i+t$ with $h_i\ge0$. The difference in \eqref{eq:increment} is
$t\DD(h,t)$, where $\DD$ is a homogeneous polynomial matrix of degree
four in $(h,t)$. We give its coefficient matrices explicitly; all are positive
semidefinite. Denote the new index by $*$, the coordinate vectors by
$e_i,e_*$, and $E_{ii}=e_ie_i^{\mathsf T}$.
For different old indices, the nonzero coefficients at $t=0$ are:
\[
\begin{array}{c|l}
\text{monomial}&\text{coefficient matrix}\\\hline
 h_i^4&E_{ii}\\
 h_i^3h_j&2(e_i-e_j-e_*)(e_i-e_j-e_*)^{\mathsf T}+E_{jj}+E_{**}\\
 h_i^2h_j^2&2(e_i-e_j)(e_i-e_j)^{\mathsf T}\\
 h_i^2h_jh_k&2(E_{jj}+E_{kk}+E_{**})
\end{array}
\]
These are manifestly positive semidefinite, and monomials in four distinct
$h$'s do not occur. The remaining coefficients
are specified below by their entries; matrices are symmetric and all
unspecified entries are zero. In the first entry table $j\ne i$ is any
other old index, so that each of these matrices has the $(m-1)$-dimensional
diagonal block $b\,I$ on those indices; in the second $i,j,k$ are distinct
old indices.
\begin{center}
\renewcommand{\arraystretch}{1.18}
\begin{tabular}{c|rrrrrr}
 & $ii$ & $jj$ & $ij$ & $i*$ & $j*$ & $**$\\\hline
 $th_i^3$ & $2m+3$ & $3$ & $-2$ & $-2m-2$ & $2$ & $3m+2$\\
 $t^2h_i^2$ & $10m+4$ & $2m+7$ & $-8$ & $-8m-8$ & $6$ & $m^2+8m+7$\\
 $t^3h_i$ & $3m^2+11m+3$ & $8m+4$ & $-10$ & $-10m-10$ & $0$ & $4m^2+8m+6$
\end{tabular}
\medskip

\begin{tabular}{c|rrrrrrr}
 & $ii$ & $jj$ & $kk$ & $ij$ & $i*$ & $j*$ & $**$\\\hline
 $th_i^2h_j$ & $12$ & $2m+9$ & $2$ & $-10$ & $-8$ & $6$ & $2m+7$\\
 $t^2h_ih_j$ & $8m+14$ & $8m+14$ & $10$ & $-18$ & $-6$ & $-6$ & $10m+8$
\end{tabular}
\end{center}
The coefficient of $th_ih_jh_k$ is diagonal, with entries $10$ at
$i,j,k$ and $12$ at $*$. The coefficient of $t^4$ has entries
\[
 Q_{ii}=3m^2+3m+1,\qquad Q_{ij}=-4\ (i\ne j),\qquad
 Q_{i*}=-4(m+1),\qquad Q_{**}=(m+1)^3.
\]
These tables follow by substituting $g_i=h_i+t$ in \eqref{eq:kernel}
and collecting monomials.

The $t^4$ and $t^2h_ih_j$ coefficients are diagonally dominant with
nonnegative diagonal: for $t^4$ the margins are $3m^2-5m+1$ in an old row
and $(m-1)^2(m+1)$ in the row $*$, and for $t^2h_ih_j$ they are $8m-10$,
$10$ and $10m-4$, all nonnegative for $m\ge2$.
For $th_i^2h_j$, eliminating the entry $12$
leaves isolated entries $2$ and the block
\[
 \frac13\begin{pmatrix}6m+2&-2\\-2&6m+5\end{pmatrix}\succeq0.
\]
For a row of the first entry table, denote its six entries by $a,b,p,q,r,c$.
The $m-1$ old indices $j\ne i$ carry the diagonal block $b\,I_{m-1}$ with
$b>0$ and couple only to $i$ and $*$; eliminating them and multiplying the
Schur complement by $b$ gives
\[
 T=\begin{pmatrix}
 ab-(m-1)p^2&bq-(m-1)pr\\
 bq-(m-1)pr&bc-(m-1)r^2
 \end{pmatrix}.
\]
Note that the original matrix is positive semidefinite if and only if
$T\succeq0$. For the middle row, we can also replace $T$ by
$\diag(6,5)T\diag(6,5)$, which is a congruence. In all three rows $T_{12}<0$, and
the two diagonal-dominance margins
$T_{11}-|T_{12}|$, $T_{22}-|T_{12}|$, with $u=m-2\ge0$, are respectively
\[
\begin{array}{c|ll}
 th_i^3&3&3u+6\\
 t^2h_i^2&6(40u^2+124u+120)&5(10u^3+79u^2+222u+9)\\
 t^3h_i&24u^3+164u^2+216u+40&32u^3+192u^2+344u+160.
\end{array}
\]
These are all nonnegative, hence each $T\succeq0$.
Thus, every coefficient of $\DD(h,t)$ is positive semidefinite. Since
$h_i,t\ge0$, this proves \eqref{eq:increment}. Starting from the case
$m=2$ verified above, ordering the $g_i$ decreasingly and adding them one
at a time so that the newly added entry is always the smallest,
proves $\K(g)\succeq0$ for every $m$.

To obtain \eqref{eq:Z}, let the $g_i$ be the combined eigenvalues of
$A,B$, and let $P_i$ be the corresponding rank-one spectral projections,
all acting on the common space. Expansion gives
$Z(A,B)=\sum_{i,j}\K(g)_{ij}P_iP_j$; products of different projections
belonging to the same matrix vanish, and the surviving cross terms
reproduce the off-diagonal part of \eqref{eq:kernel}. Factoring
$\K=R^{\mathsf T}R$ with $R$ real therefore gives
$Z(A,B)=\sum_\ell(\sum_iR_{\ell i}P_i)^2\succeq0$, each
$\sum_iR_{\ell i}P_i$ being Hermitian.
\end{proof}

\begin{lemma}\label{lem:matrix}
Suppose that\/ $\sigma>0$, $\Tr\sigma=1$, and\/ $X=\sigma^{1/2}J\sigma^{1/2}$,
where $J=J^*=J^{-1}$. If\/ $J=I$ or $\Tr J=0$, then
$\PP(X)\le\QQ_\sigma(X)$.
\end{lemma}
\begin{proof}
Since $\sigma>0$ and $J$ is invertible, so is $X$. Set
$G=|X|>0$, $S=\operatorname{sgn}X$, and
$E=G^{-1/2}\sigma G^{-1/2}>0$. The identity $X\sigma^{-1}X=\sigma$
gives $SES=E^{-1}$, so $\log_2 E$ anticommutes with $S$, whereas $G$
commutes with $S$. Since $X$ is congruent to $J$, its sign sectors have
the same dimension $k$ when $\Tr J=0$. 

In the eigenbasis of $S$, the
anticommuting matrix $\log_2 E$ is off-diagonal. A singular value
decomposition of its off-diagonal block, followed by the corresponding
change of basis inside each sector, which preserves $S$ and the block
structure of $G$, gives in a suitable orthonormal basis:
\begin{equation}\label{eq:normal}
 G=A\oplus B,\qquad S=I\oplus(-I),\qquad
 E=\begin{pmatrix}C&T\\T&C\end{pmatrix},\qquad
 C=\diag(z_i),\qquad T=\diag(s_i),
\end{equation}
where $A,B$ are positive $k\times k$ matrices, $z_i\ge1$, and
$s_i=\sqrt{z_i^2-1}\ge0$.
If $J=I$, then $X=G=\sigma$ and $E=I$.

To vary $E$ with $G$ fixed, set \[\kappa=\Tr(EG), \qquad
\sigma_E=G^{1/2}EG^{1/2}/\kappa, \qquad X_E=GS/\kappa, \qquad N_j=\Tr G^j.\]
For the original pair, we have $\kappa=1$, $\sigma_E=\sigma$, and $X_E=X$.
Substitution in \eqref{eq:P}--\eqref{eq:Q}, using $SG=GS$, $S^2=I$ and
$SES=E^{-1}$, gives
\begin{align}
 \kappa^5\bigl(\QQ_{\sigma_E}(X_E)-\PP(X_E)\bigr)
 =\FF(G,E):={}&\kappa^3N_2+2\Tr(EG^3)\Tr(EGEG)\notag\\
 &-4\kappa\Tr(EGEG^3)+2\kappa\Tr(EG^2E^{-1}G^2)
       -\kappa\PP(G).\label{eq:F}
\end{align}
At $E=I$, with $g_i$ the eigenvalues of $G$, this becomes
\begin{equation}\label{eq:F0}
 \FF(G,I)=2\sum_{i<j}g_ig_j(g_i-g_j)^2(N_1-g_i-g_j)\ge0,
\end{equation}
where each factor $N_1-g_i-g_j$ is a sum of the remaining eigenvalues.
This handles the case $J=I$. 

For \eqref{eq:normal}, we will show that
$\FF(G,E)\ge\FF(G,I)$.
Put \[c_i=(A+B)_{ii}>0, \qquad e_i=(A^3+B^3)_{ii}, \qquad N=N_2, \qquad
U_{ij}^{r,\epsilon}=(A^r)_{ij}+\epsilon(B^r)_{ij}, \qquad \epsilon=\pm1.\]
Define the real coefficients
\begin{equation}\label{eq:phi}
 \Phi_{ij}^{\ell}(\epsilon)
 =e_\ell|U_{ij}^{1,\epsilon}|^2+c_\ell|U_{ij}^{2,-\epsilon}|^2
   -2c_\ell\Rea(U_{ij}^{1,\epsilon}\overline{U_{ij}^{3,\epsilon}}),
\end{equation}
which are symmetric in $i,j$, and the linear functions
\[
 L_{ij}(z)=\tfrac12\sum_\ell z_\ell
          \bigl(\Phi_{ij}^{\ell}(1)-\Phi_{ij}^{\ell}(-1)\bigr).
\]
Choose symmetric signs $\epsilon_{ij}$ with $\epsilon_{ii}=1$, and set
\begin{align}
 H_\epsilon(z)&=N(c\cdot z)^3+
       2\sum_{i,j,\ell}z_iz_jz_\ell\Phi_{ij}^{\ell}(\epsilon_{ij}),\label{eq:H}\\
 B_\epsilon(z)&=H_\epsilon(z)-(c\cdot z)\PP(G)
                         -2\sum_{i,j}\epsilon_{ij}L_{ij}(z).\label{eq:B}
\end{align}
Block multiplication in \eqref{eq:F} yields
\begin{equation}\label{eq:comparison}
 \FF(G,E)-B_\epsilon(z)
 =2\sum_{i,j}L_{ij}(z)\bigl(s_is_j-\epsilon_{ij}(z_iz_j-1)\bigr).
\end{equation}
The diagonal terms vanish, because $s_i^2=z_i^2-1$ and $\epsilon_{ii}=1$.
Since $L_{ij}=L_{ji}$ and $|s_is_j|\le z_iz_j-1$,
the choice $\epsilon_{ij}=-\operatorname{sgn}L_{ij}(z)$ off the diagonal
makes each remaining bracket contribute
$2L_{ij}s_is_j+2|L_{ij}|(z_iz_j-1)\ge0$, so $\FF(G,E)\ge B_\epsilon(z)$;
either sign is allowed at zero.
Fix these signs below. Taking $z=\one=(1,\ldots,1)$ in
\eqref{eq:comparison}, where $s=0$ and $E=I$, gives
$B_\epsilon(\one)=\FF(G,I)$.

We will show that every coefficient of the cubic $H_\epsilon$ is
nonnegative. Set $r_i=e_i/c_i\in[0,N]$, using $c_i>0$ and
$G^3\preceq\|G\|^2G\preceq NG$.
Let $a_i,b_i$ denote the coordinate vectors of the two blocks of
$G=A\oplus B$, so that for either sign the vector $a_i+b_i$ and the
vector $a_j+\epsilon b_j$ satisfy
$\langle a_i+b_i,G^r(a_j+\epsilon b_j)\rangle=U^{r,\epsilon}_{ij}$,
$\langle a_i+b_i,G(a_i+b_i)\rangle=c_i$ and
$\langle a_i+b_i,G^3(a_i+b_i)\rangle=e_i$.
Cauchy--Schwarz for the forms $\langle\cdot,G\,\cdot\rangle$ and
$\langle\cdot,G^3\cdot\rangle$ therefore gives
\[
 v:=\frac{|U_{ij}^{1,\epsilon}|}{\sqrt{c_ic_j}}\le1,
 \qquad |U^{3,\epsilon}_{ij}|\le\sqrt{e_ie_j},
 \qquad\text{hence}\qquad
 \frac{\Rea(U_{ij}^{1,\epsilon}\overline{U_{ij}^{3,\epsilon}})}{c_ic_j}
 \le\sqrt{r_ir_j}\,v.
\]
Applying \eqref{eq:sharp} to $G=A\oplus B$ and the vectors
$a_i+b_i$, $a_j-\epsilon b_j$, whose $G^2$ pairing is
$U^{2,-\epsilon}_{ij}$ and whose $G$- and $G^3$-norms are again
$c_i,c_j$ and $e_i,e_j$, leads to
\[
 \frac{|U_{ij}^{2,-\epsilon}|^2}{c_ic_j}\ge(r_i+r_j-N)_+ .
 \]
Therefore,
\begin{align*}
 \frac{\Phi_{ij}^{\ell}(\epsilon)}{c_ic_jc_\ell}
 &\ge r_\ell v^2-2\sqrt{r_ir_j}\,v+(r_i+r_j-N)_+
\\
 &\ge(r_i+r_j-N)_+-h(r_i,r_j;r_\ell).
\end{align*}
These inequalities also hold for repeated indices. Since the three
products $c_ic_jc_k$, $c_jc_kc_i$, $c_kc_ic_j$ coincide,
\eqref{eq:three} with $N=N_2$ proves that
\[
 \tfrac32Nc_ic_jc_k+
 \Phi_{ij}^{k}(\epsilon_{ij})+
 \Phi_{jk}^{i}(\epsilon_{jk})+
 \Phi_{ki}^{j}(\epsilon_{ki})\ge0,
\]
for every $i,j,k$. Multiplying by $\tfrac23z_iz_jz_k$ and summing over all ordered triples
gives exactly $H_\epsilon(z)$, since each of the three $\Phi$ terms sums
to $\sum_{i,j,k}z_iz_jz_k\Phi^k_{ij}(\epsilon_{ij})$. This proves the
coefficient claim without separating cases.

It remains to check the gradient at $\one$. Differentiating
\eqref{eq:B} gives
\begin{equation}\label{eq:gradient}
 \tfrac12\partial_i B_\epsilon(\one)=Z(A,V_iBV_i)_{ii}\ge0,
 \qquad V_i=\diag(\epsilon_{i1},\ldots,\epsilon_{ik}),
\end{equation}
using that $V_iBV_i$ is positive semidefinite by Lemma~\ref{lem:operator}.
To verify the identity, its left side expands as
\[
 \Gamma c_i+N_2e_i+2N_3(A^2+B^2)_{ii}-2N_1(A^4+B^4)_{ii}
       +2\sum_j\epsilon_{ij}L_{ij}(\one),
\]
which is the diagonal of \eqref{eq:Z} after conjugating $B$ by $V_i$.
Indeed, $(A+V_iBV_i)_{ij}=U^{1,\epsilon_{ij}}_{ij}$, and similarly for the
higher powers, and collecting the terms of $Z(A,V_iBV_i)_{ii}$ that carry
a factor $\epsilon_{ij}$ reproduces $2L_{ij}(\one)$.
Since $B_\epsilon-H_\epsilon$ is linear and $H_\epsilon$ has nonnegative coefficients, each partial derivative of $B_\epsilon$ is coordinate-wise non-decreasing
on the nonnegative orthant, so \eqref{eq:gradient} gives
$\partial_iB_\epsilon(z)\ge\partial_iB_\epsilon(\one)\ge0$ for $z\ge\one$, and
hence $B_\epsilon(z)\ge B_\epsilon(\one)$ there. As $z\ge\one$ in
\eqref{eq:normal}, we get
$\FF(G,E)\ge B_\epsilon(z)\ge B_\epsilon(\one)=\FF(G,I)\ge0$,
and \eqref{eq:F} with $\kappa=1$ proves the lemma.
\end{proof}

\begin{proof}[Proof of Theorem~\ref{thm:main}]
Padding the smaller register with a product of $|0\rangle$ qubits changes
neither $F$ nor $W$: it leaves the nonzero spectrum of $\rho_A$ alone,
and it doubles both $d_Ad_B$ and $\sum_{P,Q}\langle P\otimes Q\rangle^4$,
because $\langle0|P|0\rangle$ vanishes for $P\in\{X,Y\}$ and
equals $1$ for $P\in\{\mathbb{I},Z\}$.
Thus, without loss of generality, we can assume that $d_A=d_B=d$. 

Let us write
$\psi=\sum_i|i\rangle\otimes T|i\rangle$ with a $d\times d$ matrix $T$ such that $\Tr (TT^*)=1$.
Invertible normalized coefficient matrices are dense in the set
$\{T:\Tr (TT^*)=1\}$, and $4F\le1-W$ is a closed condition, so by continuity
of $F$ and $W$ it suffices to treat invertible $T$.
Put $\sigma=TT^*>0$ and $X_P=TP^{\mathsf T}T^*$. Then
$\langle\psi|P\otimes Q|\psi\rangle=\Tr(X_PQ)$, and $\rho_A=(T^*T)^{\mathsf T}$
has the same spectrum as $\sigma$.
The polar decomposition $T=\sigma^{1/2}U$ expresses $X_P$ as
$\sigma^{1/2}J_P\sigma^{1/2}$, where $J_P=UP^{\mathsf T}U^*$
is a Hermitian involution. It is traceless unless $P=I$, in which case
$J_I=I$, so
Lemmas~\ref{lem:pauli} and \ref{lem:matrix} apply to every Pauli row.
Pauli twirling, $\sum_PPYP=d\,\Tr(Y)\,I$, gives for every matrix $Y$:
\[
 \frac1d\sum_PX_PYX_P=\Tr(\sigma Y)\,\sigma .
\]
Averaging \eqref{eq:Q} over $P$ and using this identity term by term, with
$Y=I$ in $\Tr X_P^2$ and $\Tr(\sigma X_P^2)$ and $\Tr(\sigma^2X_P^2)$, and
with $Y=\sigma$ in $\Tr(\sigma X_P\sigma X_P)$, gives
\[\frac1d\sum_P\QQ_\sigma(X_P)=1+4(\Tr\sigma^2)^2-4\Tr\sigma^3.\] Hence
\[
 W=\frac1{d^2}\sum_P\sum_Q\Tr(X_PQ)^4
   \le\frac1d\sum_P\PP(X_P)
   \le\frac1d\sum_P\QQ_\sigma(X_P)
   =1-4\bigl(\Tr\sigma^3-(\Tr\sigma^2)^2\bigr)=1-4F,
\]
the first inequality being Lemma~\ref{lem:pauli} applied to each $X_P$ and
summed over $P$, and the second Lemma~\ref{lem:matrix}.

To see that the coefficient $4$ is optimal, consider
$\psi=\sqrt p\,|00\rangle+\sqrt{1-p}\,|11\rangle$, and put
$s=p(1-p)$. A direct calculation gives $F=s-4s^2$ and $W=1-4s+16s^2$.
Thus, equality holds with $F>0$, whenever $0<p<1$ and $p\ne1/2$.
\end{proof}

\end{document}